%% file: prep_ml.tex
\documentclass[]{article}
\usepackage{multicol}
\usepackage{apj}
\def\lesssim{\mathrel{\hbox{\rlap{\hbox{\lower4pt\hbox{$\sim$}}}\hbox{$<$}}}}
\def\gtrsim{\mathrel{\hbox{\rlap{\hbox{\lower4pt\hbox{$\sim$}}}\hbox{$>$}}}}
\newcommand{\mincir}{\raise -2.truept\hbox{\rlap{\hbox{$\sim$}}\raise5.truept
\hbox{$<$}\ }}
\newcommand{\magcir}{\raise -2.truept\hbox{\rlap{\hbox{$\sim$}}\raise5.truept
\hbox{$>$}\ }}
\newcommand{\siml}{\raise -2.truept\hbox{\rlap{\hbox{$\sim$}}\raise5.truept
\hbox{$<$}\ }}
\newcommand{\simg}{\raise -2.truept\hbox{\rlap{\hbox{$\sim$}}\raise5.truept
\hbox{$>$}\ }}
\newcommand{\be}{\begin{equation}}
\newcommand{\ee}{\end{equation}}
\newcommand{\ba}{\begin{eqnarray}}
\newcommand{\ea}{\end{eqnarray}}
\newcommand {\h} {$h^{-1}$ Mpc $ \;$}
\newcommand {\hh} {$h^{-1}$ Mpc}
\newcommand {\ks} {km~s$^{-1} \;$}

\newcommand {\ml} {$h \, M_{\odot}/L_{\odot} \;$}

\begin{document}


\vspace{15mm}                                                                   
\begin{center}
\uppercase{
Optical Luminosities and Mass--to--Light Ratios \\ of Nearby Galaxy 
Clusters}\\
\vspace*{1.5ex}
{\sc Marisa Girardi$^{1}$, Stefano Borgani$^{2}$,
Giuliano Giuricin$^{1,3}$,
Fabio Mardirossian$^{1,4}$,\\
and Marino Mezzetti$^{1}$}\\
\vspace*{1.ex}
{\small
$^1$ 
Dipartimento di Astronomia, Universit\`{a} 
degli Studi di Trieste, Via Tiepolo 11, I-34131 Trieste, Italy\\
$^2$ INFN, Sezione di Perugia,  c/o Dipartimento di Fisica
dell'Universit\`a, via A. Pascoli, I-06123 Perugia, Italy;\\
INFN, Sezione di Trieste, c/o 
Dipartimento di Astronomia, Universit\`{a} 
degli Studi di Trieste, Via Tiepolo 11, I-34131 Trieste, Italy\\
$^3$SISSA, via Beirut 4, I-34014 Trieste, Italy\\
$^4$Osservatorio Astronomico di Trieste, Via Tiepolo 11, I-34131 Trieste, 
Italy\\
E-mail: girardi, borgani, giuricin, mardiros, mezzetti @ts.astro.it.it \\}
\end{center}
\vspace*{-6pt}

\begin{abstract}

We analyze a sample of 105 clusters having virial mass homogeneously
estimated and for which galaxy magnitudes are available with a well
defined high degree of completeness.  In particular, we consider a
subsample of 89 clusters with $B_j$ band galaxy magnitudes taken from
the COSMOS/UKST Southern Sky Object Catalogue.

After suitable magnitude corrections and uniform conversions to $B_j$
band, we compute cluster luminosities $L_{B_j}$ within several
clustercentric distances, $0.5,1.0,1.5$ \h and within the
virialization radius $R_{vir}$. In particular, we use the luminosity
function and background counts estimated by Lumsden et al. (1997) on
the Edinburgh/Durham Southern Galaxy Catalogue, which is the
well--calibrated part of the COSMOS catalogue. We analyze the effect
of several uncertainties connected to photometric data,
fore/background removal, and extrapolation below the completeness
limit of the photometry, in order to assess the robustness of our
cluster luminosity estimates.

We draw our results on the relations between luminosity and dynamical
quantities from the COSMOS sample by considering mass and luminosities
determined within the virialization radius.  We find a very good
correlation between cluster luminosity, $L_{B_j}$, and galaxy velocity
dispersion, $\sigma_v$, with $L_{B_j} \propto \sigma_v^{\mbox{\rm
2.1--2.3}}$.  Our estimate of typical value for the mass-to-light
ratio is $M/L_{B_j}\sim 250$ \ml.  We do not find any correlation of
$M/L_{B_j}$ with cluster morphologies, i.e. Rood--Sastry and
Bautz--Morgan types, and only a weak significant correlation with
cluster richness.

We find that mass has a slight, but significant, tendency to increase
faster than the luminosity does, $M\propto L_{B_j}^{\mbox{\rm
1.2--1.3}}$.  We verify the robustness of this relation against a
number of possible systematics. We verify that this increasing trend
of $M/L$ with cluster mass cannot be entirely due to a higher spiral
fraction in poorer clusters, thus suggesting that a similar
result would also be found by using
$R$ band galaxy magnitudes.


%
\vspace*{6pt}
\noindent
{\em Subject headings: }
galaxies: clusters: general - galaxies: fundamental parameters --
cosmology: observations.

\end{abstract}

\begin{multicols}{2} 

\section{INTRODUCTION}

Since the work by Zwicky (1933), it is well known that the luminous
matter associated to galaxies in clusters provides only a small part
of the total cluster mass. Pioneering works typically found
mass--to--optical-luminosity ratios, $M/L$, of several hundreds in
solar units. For instance Faber \& Gallagher (1979) computed $M/L$ for
seven clusters and reported a median value of $M/L_V \sim 580$ \ml,
where $L_V$ is the $V$ band luminosity, corresponding to $M/L_B \sim
750$ \ml when using $B$ band luminosity (hereafter $h$ is the Hubble
constant in units of 100 \ks $Mpc^{-1}$).  Dressler 
(1978b) found $M/L_R \sim 200$--$600$ \ml for eight clusters
(corresponding to about $M/L_B \sim 300$--$900$ \ml).  Subsequent
analyses of mass estimates based on optical and X--ray data indicated
smaller values: 
$M/L_B=190$ \ml for A194 by Chapman, Geller, \& Huchra
(1988); 
$M/L_B=200$ \ml for Perseus cluster by Eyles et
al. (1991); 
$M/L_V \sim 200$--$300$ \ml for seven groups/clusters by
David, Jones, \& Forman (1995); $M/L_{B}\sim 400$ for 29 clusters of
ENACS (ESO Nearby Abell Clusters Survey, Adami et al. 1998b).  Small
values are found also in the case of more distant clusters $M/L_R \sim
300 $ \ml (Carlberg et al. 1996 for clusters of the Canadian Network
for Observational Cosmology, CNOC).  However, high values of $M/L$
have been still recently found by Mohr et al. (1996), who give
$M/L_R=760$--$1600$ \ml for A576 from optical mass estimates.

According to the common wisdom, $M/L$ increases from galaxy to cluster
scales (e.g. Blumenthal et al. 1984), and saturates at the level of
galaxy systems (see also Rubin 1993; Bahcall, Lubin, and Dorman
1995). The analysis by Rubin (1993)  suggested that all
systems have a constant ratio of $M/L_B\sim 200$ \ml for scales larger
than galaxies, $\sim 50 h^{-1} kpc$, so that the total mass of galaxy
systems could be roughly accounted for by the total mass of their
member galaxies (see also Bahcall et al. 1995).

The above results are based on $M/L$ estimates for individual clusters
or for small cluster samples.  Indeed, observational difficulties
prevented from building a large $M/L$ data base, where both masses and
luminosities are computed in a homogeneous way.  The determination of
cluster luminosities is fraught with several uncertainties, related to
corrections for Galactic extinction and background galaxy
contamination, calibration of the photometry, and correction of (the
usual) isophotal galaxy magnitudes to total magnitudes. Further
complications arise due to the need of extrapolating the sum of the
measured luminosities of the cluster members to include galaxies below
the completeness limit of the photometry, and to include the outer
parts of the cluster beyond the region studied.  These difficulties
limited the number of clusters with homogeneous luminosity
determinations (e.g. Oemler 1974 -- 15 clusters; Dressler 1978a -- 12
clusters; Adami et al. 1998a -- 29 clusters; 
Carlberg et al. 1996 -- 16
distant clusters).

Also the estimate of cluster masses is not an easy task in spite of
the various methods which are applied.  Cluster masses are inferred
from either $X$--ray or optical data, under the general hypothesis of
dynamical equilibrium.  Estimates based on gravitational lensing do
not require assumptions about the dynamical status of the cluster, but
a good knowledge of the geometry of the potential well is necessary
(e.g., Narayan \& Bartelmann 1996).  Claims for a discrepancy (by a
factor of 2--3) between cluster masses obtained with different methods
casted doubts about the general reliability of mass estimates (e.g.,
Wu \& Fang 1997).  However, recent analyses have shown that such
discrepancies can be explained by the different way in which strong
cluster substructures bias mass estimates based on different methods
(Allen 1997; Girardi et al. 1997). In particular, Girardi et
al. (1998b, hereafter G98) showed that, for nearby clusters without
strong substructures, a good overall agreement exists between $X$--ray
and optical mass estimates. A similar result was obtained by Lewis
et al. (1999) for the distant clusters of CNOC. 

The sample of 170 nearby clusters analyzed by G98 is the largest one
available with homogeneous mass estimates. Therefore, this sample is
suitable for $M/L$ determinations once a proper analysis of cluster
luminosities is made.  To this purpose, we resort to the large
homogeneous data base for $B_j$ galaxy photometry provided by the
COSMOS/UKST Southern Sky Object Catalogue (Yentis et al. 1992). We use
the results from the careful work by Lumsden et al. (1997, hereafter
L97) on the well-calibrated part of the COSMOS catalogue known as the
Edinburgh/Durham Southern Galaxy Catalogue (EDSGC, Heydon-Dumbleton,
Collins, \& MacGillivray 1989). In particular, we use their best fit
luminosity function and their counts of background galaxies for
estimating cluster luminosities.

The paper is organized as follows.  We describe the data sample and
compute cluster luminosities in \S~2 and \S~3, respectively.  The
resulting luminosities and relative uncertainties are analyzed in
\S~4. We compute mass--to--light ratios in \S~5.  We devote \S~6 to
the analysis of the relations between luminosity and dynamical quantities.
We discuss our results in \S~7, while we give
in \S~8 a brief summary of our main results and draw our conclusions.

\section{THE DATA SAMPLE}

From the sample of nearby clusters ($z\le 0.15$) analyzed by G98 we
draw a subsample of 105 clusters for which galaxy magnitudes are
available.  In particular, we avoid the G98 clusters which show two
peaks either in the velocity or in the projected galaxy distribution,
as well as clusters with uncertain dynamics (cf. \S~2 of G98). We also
exclude A3562 which overlaps with SC1329-314 and A754, which is
well--known for having a bimodal structure (e.g. Zabludoff \& Zaritsky
1995).  From the G98 analysis we take for each cluster: the
line--of--sight velocity dispersion $\sigma_v$, the radius of
virialization $R_{vir}$, and the virial mass $M$ within $R_{vir}$. For
an appropriate comparison between masses and luminosities, the latter
are computed within regions centered on the same cluster centers used
by G98.

\subsection{The COSMOS Sample}

We draw most of our results from a homogeneous sample of 89 clusters
("$C$--sample" hereafter) for which galaxy $B_j$ magnitudes and
positions are available from the COSMOS catalogue (Yentis et
al. 1992); 20 out of these 89 clusters are in the EDSGC
(Heydon-Dumbleton et al. 1989).

The EDSGC is nominally quasi--complete to $B_j=20$. A more
conservative limiting apparent magnitude was suggested by Valotto et
al. (1997) who found that galaxy counts follow a uniform law for
$B_j<19.4$. Accordingly, we decide to adopt a limiting magnitude of
$B_j=19.4$ for the entire COSMOS catalog.  By using data of the ESO
Nearby Abell Cluster Survey (ENACS), Katgert et al. (1998) assessed
that COSMOS is $91\%$ complete and suggested that this completeness is
more appropriate for areas with high surface density, rather than the
nominal $95\%$ of the catalog (Heydon-Dumbleton et al. 1989). We adopt
a value of $91\%$ for the cluster completeness in the $C$--sample.

For each cluster we select circular regions with a radius taken to be
the largest between $R_{vir}$ and the Abell radius (1.5 \hh). For some
clusters the data sample is obtained by joining data from more than
one plate, since both magnitude and position inter--plate variations
are small (Heydon-Dumbleton et al. 1989; Drinkwater, Barnes, and
Ellison 1995).

We correct each galaxy magnitude for (1)~Galactic absorption by
assuming the absorption in blue band as given by de Vaucouleurs et
al. (1991), and (2) $K$--dimming by assuming that all galaxies lie at
the average cluster redshift (Colless 1989).  Moreover, we apply the
correction of L97 to convert COSMOS $B_j$ magnitudes to the CCD based
magnitude scale. This correction is consistent with our choice to use
the luminosity function and counts by L97.

Finally, we note that COSMOS magnitudes are isophotal magnitudes (the
threshold is on average only $8\%$ above the sky,
cf. Heydon-Dumbleton, Collins, \& MacGillivray 1988).  However,
Shanks, Stevenson, \& Fong (1984) argue that only at faint limit (well
below our limiting magnitude) the difference between COSMOS magnitudes
and ``total'' magnitudes becomes significant.

\subsection{Other Samples}

In addition to the $C$--sample, we also analyze two cluster samples
which are not homogeneus in their photometry in order to enlarge the
list of available cluster luminosities as well as to perform a more
accurate analysis of the errors associated to the computation of
luminosities.

The second sample contains data for 39 clusters which have published
galaxy positions and magnitudes down to a given completeness magnitude
and are sampled at least out to 0.5 \h ("$M$--sample", hereafter).
This sample is inhomogeneous in photometry since it contains 25, 2,
and 12 clusters with photometry in blue, visual, and red bands,
respectively.  

A third sample (``$Z$--sample'', hereafter) includes those 17 clusters
of the $M$--sample, for which galaxy redshifts are also available
within a certain level of completeness.  We include in this sample
also Ursa Major and A3558, thus ending up with 19 clusters in the
$Z$--sample. The Ursa Major cluster data have galaxy redshifts of only
cluster members up to a given completeness magnitude limit. For A3558
we use the redshift compilation by Bardelli et al. (1998, which now
also include ENACS data) where magnitudes are taken from the COSMOS
catalog, with a resulting completeness of $97\%$ for $B_j<18$ within
$\sim 0.75$ \hh.

For clusters of the $M$-- and $Z$--samples we adopt the limiting
magnitude and the level of completeness provided by the authors of
data within the respective region of completeness.  For clusters of
the $Z$--sample we consider only galaxies identified as cluster
members: for cluster data in common with G98 we use their member
identification after the peak selection in velocity distribution via
the adaptive kernel method and the shifting gapper (see also Fadda et
al. 1996). For new cluster data we apply this same procedure for
cluster member identification.  The $M$--sample is generally superior
to the $Z$--sample in the photometry depth (the median limiting
absolute magnitude of $M$--clusters is $M_{B_j}=-17.6$ comparable to
that of $C$--clusters, and only $M_{B_j}=-18.4$ for $Z$--clusters).
The advantage of considering the $Z$--sample is that no
fore/background subtraction is needed to estimate luminosities (\S~3).

Table~1 lists the properties of the $M$-- and $Z$--samples,
respectively: the source of photometric data (Col.~2); the original
magnitude--band and the respective limiting apparent magnitude
(Cols.~3 and 4); the completeness (Col.~5), defined as the fraction of
galaxies brighter than the magnitude limit used in our analysis.  For the
$M$--samples this fraction is given by all the galaxies with measured
magnitues, while for the $Z$--sample is given by the galaxies for
which both magnitudes and redshifts are available.

\vspace{6mm} 
\hspace{-4mm}
\begin{minipage}{9cm}
\renewcommand{\arraystretch}{1.2}
\renewcommand{\tabcolsep}{1.2mm}
\begin{center}
\vspace{-3mm} 
TABLE 1\\
\vspace{2mm}
{\sc Cluster Samples\\}
\footnotesize
\vspace{2mm} 
\input{tab1a}
\end{center}
\end{minipage}

\vspace{6mm} 
\hspace{-4mm}
\begin{minipage}{9cm}
\renewcommand{\arraystretch}{1.2}
\renewcommand{\tabcolsep}{1.2mm}
\begin{center}
\vspace{-3mm} 
TABLE 1\\
\vspace{2mm}
{\sc continued\\}
\footnotesize
\vspace{2mm} 
\input{tab1b}
\end{center}
\input{comm_tab1}
\vspace{3mm}  
\end{minipage}

We correct each galaxy magnitude for: (1)~galactic absorption by
assuming the absorption in blue band as given by de Vaucouleurs et
al. (1991) and transformed to visual and red bands according to
Sandage (1973); (2) $K$--dimming by assuming that all galaxies lie at
the average cluster redshift.  We adopt the $K$--dimming corrections
given by Colless (1989) for blue band, and by Sandage (1973) for 
visual and red bands.

Then, in order to homogeneize the photometry, we convert all
magnitudes into the $B_j$ band.  In particular, since magnitudes
measured by APM and COSMOS machines suffer from similar systematic
effects (Metcalfe, Fong \& Shanks 1995), we apply the same correction,
already applied to COSMOS magnitudes, to APM magnitudes in order to
convert $B_j$ magnitudes to the CCD based magnitude scale
(cf. \S~2.1).  We transform $B$ band magnitudes into the $B_j$ band by
the expression $B_j=B+0.28(B-V)$ (Blair \& Gilmore 1982; see also
Metcalfe et al. 1995, and Maddox, Efstathiou, \& Sutherland
1990). Magnitudes in other bands are converted to the $B$ and then to
$B_j$ band, according to the following procedure: by using eqs.~(6)
and~(7) by Kirshner, Oemler, \& Schechter (1978) for $B_{Zwicky}$; by
using $B=V+(B-V)$ for the $V$ band; by using $B=R+(B-V)+(V-R)$ for the
$R$ band or the $F$ band (assuming $R=F$ as in Lugger 1989; see also
Trevese, Cirimele, \& Appodia 1996).  In the above conversions we use
the colors for the appropriate galaxy morphological type (Buta et
al. 1994; Buta \& Williams 1995) or, when morphology is unknown (i.e.
for all $C$--clusters and for most of the other clusters), the typical
colors of early type galaxies $(B-V)=0.9$ and $(V-R)=0.55$.
Magnitudes in the $r$ band by Gunn are reported to $R$ band with
$r=R+0.35$ (J\o rgensen 1994, see also Geller et al.  1997) and then
to $B$ and to $B_j$ bands.  In general, the magnitude bands are
isophotal and we do not correct for this effect.

\section{COMPUTING THE CLUSTER LUMINOSITIES}

Observed luminosities are computed within several clustercentric
regions defined by the virialization radius, $R_{vir}$, and by the fixed
apertures 0.5, 1, 1.5 \hh.  We always avoid extrapolating the
luminosity to include the outer parts of the clusters beyond the
completeness region.  Therefore, for $C$--clusters we can compute
luminosities in all the above cluster regions, while for $M$-- and
$Z$--clusters we can always compute luminosities within 0.5 \hh, but
only in some cases within more extended regions.

Observed cluster luminosities, $L_{B_j,obs}$, are obtained by summing
the individual absolute luminosities of all galaxies and assuming
$B_{j,\odot}=5.33$ (i.e.  $B_{\odot}=5.48$ and using the conversion to
$B_j$ by Kron 1978).

The resulting absolute luminosities and the number counts are
corrected for partial sampling incompleteness of the original samples
by dividing them by the nominal completeness, as provided in the
literature (see Table~1). This amounts to assume that the correction
for incompleteness is the same for counts and luminosities. This
assumption is certainly appropriate for $C$--clusters, since Katgert
et al. (1998) already found that the magnitude distribution of missing
galaxies is essentially the same as for sampled galaxies.

Luminosities of $C$-- and $M$--clusters need to be corrected for
fore/background contamination.  We compute the corrected luminosity,
$L_{B_j,corr}$, by subtracting the average fore/background luminosity
obtained from the mean field $B_j$ counts of L97 (cf. their Figure~2)
and assuming a $95\%$ completeness for the COSMOS catalog
(Heydon-Dumbleton et al. 1989). Corrected counts $N_{corr}$ are then
obtained in a similar way.  The limitation of this procedure is that
local fluctuations of the luminosity field are not taken into account.

As an alternative, we also consider the method recently used by Adami
et al. (1998a) in the analysis of the fundamental plane of ENACS
clusters. This method is based on applying suitable corrections,
$L_{B_j,corr}=L_{B_j,obs}*f_L$ and $N_{corr}=N_{obs}*f_N$, under the
assumption that the fraction of the luminosity, $f_L$, and the
fraction of the number, $f_N$, of cluster galaxies within the selected
fields in the present samples are the same as in the G98 samples.  The
G98 samples were collected by requiring that redshifts were available
for all galaxies, thus allowing the definition of galaxy membership.
We adopt $f_L=f_N$ for those G98 clusters for which magnitudes are not
available.

The median fore/background correction for $C$-- and $M$--clusters lies
in the range $\sim 15$--$40\%$ depending on the method and on the
extension of the cluster area (in particular it is $20$--$30\%$ within
$R_{vir}$).

In order to obtain the total cluster luminosity, we need to
extrapolate the above luminosities to include galaxies below the
magnitude completeness limit. We adopt the usual Schechter (1976) form
for the cluster luminosity function, hereafter LF, to obtain the total
cluster luminosity:
\begin{equation} 
L_{B_j,tot}=L_{B_j,corr}+\Phi^* L^*\int^{L_{lim}/L^*}_{L_{min}/L^*}
x^{1+\alpha} e^{-x}dx,
\end{equation} 
\noindent 
where $L_{B_j,corr}$ is the observed cluster luminosity
corrected for incompleteness and fore/background subtraction, $L_{lim}$
is the luminosity corresponding to the limiting magnitude, $L_{min}$
corresponds to a cut--off for the minimum galaxy luminosity (here we
adopt $L_{min}=10^{-4}L^*$), and $L^*$, $\alpha$, and $\Phi^*$ are the
parameters of the LF.  
We adopt the LF parameters determined by L97, i.e.  the $L^*$ value
corresponding to the absolute magnitude $M_{B_j}^*=-20.16$ and
$\alpha=-1.22$. L97 derived their best fit of LF over the range $-18$
to $-21$ in $M_{B_j}$, for a composite sample of 22 clusters.
The $\Phi^*$ parameter is determined from the observed (corrected)
galaxy number counts for $-21 \le M_{B_j} \le -18$,
$N_{corr}(-21,-18)$:
\begin{equation}
\Phi^*=N_{corr}(-21,-18)/\int_{L(-18)/L^*}^{L(-21)/L^*}x^{\alpha}e^{-x}dx,
\end{equation} 
\noindent where $L(-18)$ and $L(-21)$ are the luminosities
corresponding to absolute magnitudes of $M_{B_j}=-18$ and
$M_{B_j}=-21$, respectively.  If the absolute limiting magnitude is
brighter than $M_{B_j}=-18$, we take $L_{lim}$ for the lower
integration limit in eq.~(2).  Due to the extrapolation to faint
magnitudes, the luminosity typically increases by $\sim 15$--$20\%$
for $C$-- and $M$--clusters and more than twice for $Z$--clusters.

The final luminosity estimates $L_{B_j}$ ($L_{B_j,c}$ or $L_{B_j,f}$
if the L97 counts or fractions are used for the fore/background
rejection, respectively) take into account the internal galactic
absorption by adopting a correction of $\Delta B_j=0.1$ mag.  Due to
this correction, the luminosity increases by about $10\%$.

Table~2 lists the cluster luminosities: the sample to which the
cluster belongs (Col.~2); the adopted cluster center (Col.~3); the
number of galaxies and the cluster luminosities calculated within 0.5,
1, and 1.5 \h (Cols.~4--9); the virialization radius $R_{vir}$
(Col.~10), the number of galaxies, $N(R_{vir})$, and the cluster
luminosities, $L_{B_j}(R_{vir})$, calculated within $R_{vir}$
(Cols.~11 and 12). We list both $L_{B_j,c}$ and $L_{B_j,f}$ within
0.5, 1 \hh, and $R_{vir}$, while we list only $L_{B_j,c}$ within 1.5
\h~(see \S~4.1 below).

\section{ROBUSTNESS OF LUMINOSITY ESTIMATES}

Our determination of cluster luminosities is fraught with
uncertainties connected to original photometric data (quality and
calibration of the photometry, and completeness estimate), with
magnitude corrections and conversions, as well as with uncertainties
connected with the corrections applied to observed luminosities
(i.e. fore/background subtraction and extrapolation to faint
magnitudes). In the following we analyse possible random and
systematics errors, by devoting particular care to $C$--clusters which
will be used to obtain our main results (cf. \S~6).

\subsection{Estimate of fore/background Contributions}

In our analysis we consider two different ways of estimating
fore/background corrections in the computation of the $L_{B_j,c}$ and
$L_{B_j,f}$, respectively.

As for $L_{B_j,c}$, the most direct approach would be using the L97
counts for fore/background removal. However, this method does not take
into account the local field-to-field count variations, which lead to
random errors. As a possible remedy, one can estimate the background
for each cluster by using counts within distant external annuli. On
the other hand, Rauzy, Adami, \& Mazure (1998) pointed out that
changes in the local field are so strong as to make this method not
free of intrinsic biases (e.g., due to the presence of a nearby group
along the cluster line-of-sight).

\includegraphics{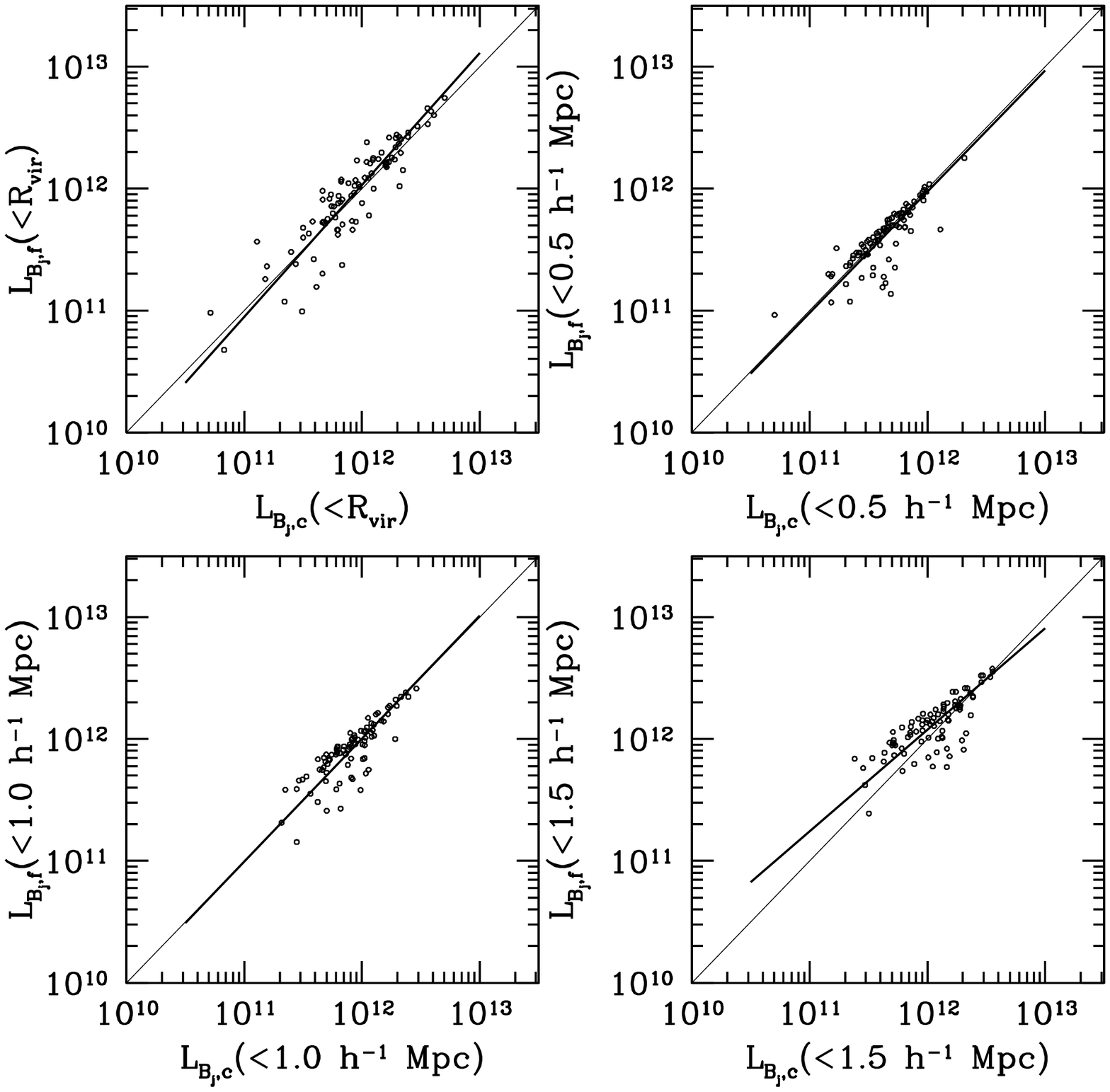}
$\ \ \ \ \ \ $\\
\vspace{9.4truecm}
$\ \ \ $\\
{\small\parindent=3.5mm {Fig.}~1.---
For $C$-- clusters we
show the comparison between $L_{B_j,c}$ and $L_{B_j,f}$, two 
alternative luminosity estimates
differing for the method of fore/background correction (see text),
as computed within several cluster regions. 
Luminosities are in units of $h^{-2}~L_{\odot}$.
Heavy lines represent the linear fits.
}

\vspace{5mm}

It is clear that having redshifts for all galaxies would allow an
unambiguous determination of the cluster membership.  By using the G98
sample of galaxy redshifts for a list of clusters, we estimate for
each cluster the luminosity fraction, $f_L$, of galaxies which are
recognized as genuine members. This fraction is then used to correct
the observed luminosity. However, $f_L$ depends on the magnitude limit
and on the extension of the sampled region around the cluster. Since
the G98 redshift samples are in general shallower and less
spatially extended than the magnitude samples we are dealing with,
this could introduce systematic errors, thus leading to an
$\ $ overestimate of $\ $
the true
\end{multicols}

\vspace{6mm} 
\hspace{-4mm}
\begin{minipage}{9cm}
\renewcommand{\arraystretch}{1.2}
\renewcommand{\tabcolsep}{1.2mm}
\begin{center}
\vspace{-3mm} 
TABLE 2\\
\vspace{2mm}
{\sc Luminosity Estimates\\}
\footnotesize
\vspace{2mm} 
\input{tab2a}
\end{center}
\end{minipage}

\vspace{6mm} 
\hspace{-4mm}
\begin{minipage}{9cm}
\renewcommand{\arraystretch}{1.2}
\renewcommand{\tabcolsep}{1.2mm}
\begin{center}
\vspace{-3mm} 
TABLE 2\\
\vspace{2mm}
{\sc continued\\}
\footnotesize
\vspace{2mm} 
\input{tab2b}
\end{center}
\end{minipage}

\vspace{6mm} 
\hspace{-4mm}
\begin{minipage}{9cm}
\renewcommand{\arraystretch}{1.2}
\renewcommand{\tabcolsep}{1.2mm}
\begin{center}
\vspace{-3mm} 
TABLE 2\\
\vspace{2mm}
{\sc continued\\}
\footnotesize
\vspace{2mm} 
\input{tab2c}
\end{center}
\end{minipage}

\begin{multicols}{2} 

\vspace{6mm} 
\hspace{-4mm}
\begin{minipage}{9cm}
\renewcommand{\arraystretch}{1.2}
\renewcommand{\tabcolsep}{1.2mm}
\begin{center}
\vspace{-3mm} 
\footnotesize
\end{center}
\input{comm_tab2}
\vspace{3mm}  
\end{minipage}

\noindent   $L_{B_j,f}$.
In addition to this systematic
uncertainty, there is also a random uncertainty due to the fact that
the G98 sample has neither a homogeneous magnitude limit nor a
uniform spatial extension around each cluster.

In order to estimate by how much such uncertainties affect our
estimates of cluster luminosities, we compare in Figure~1 $L_{B_j,c}$
and $L_{B_jf}$ for $C$--clusters within different radii. It is
apparent that these two alternative estimates are in overall
agreement, although within some scatter.  While the scatter provides
an estimate of the random errors, the agreement indicates that any
sistematic bias should not seriously pollute our analysis. For
instance, for $C$--clusters we find a scatter of $\sim 30\%$ and
$L_{B_j,c}/L_{B_j,f}=0.88$ for luminosities estimated within
$R_{vir}$.  Somewhat different results hold for luminosities within
1.5 \h (cf. lower--right panel in Fig. 1). In this case, the
discrepancy is due to the fact that redshift data generally sample
clusters only out to their typical size. This turns into high,
unreliable $L_{B_j,f}$ determinations when poor clusters are examined
within the large region encompassed by the 1.5\h radius. For this
reason we do not report in Table~2 $L_{B_j,f}$ luminosities computed
within this radius.

Another way of assessing the impact of fore/background corrections is
based on comparing the luminosities of the $Z$--clusters, which are
free from these corrections (cf. \S2.2), with those of the
corresponding $M$--clusters (or $C$--cluster for A3558), which have
the same original photometry. For this comparison we use luminosities
within 0.5 \hh, so as to maximize the number of clusters (18) which
belong to both samples.  Figure~2 shows that there is no appreciable
systematic effect and that the scatter is about $10\%$. Allowing for
larger and deeper data samples, it is reasonable to expect that the
contribution of fore/background corrections to random errors increases
to $\sim 20\%$, thus comparable to the above estimate of the scatter
(cf. Fig. 1).

\includegraphics{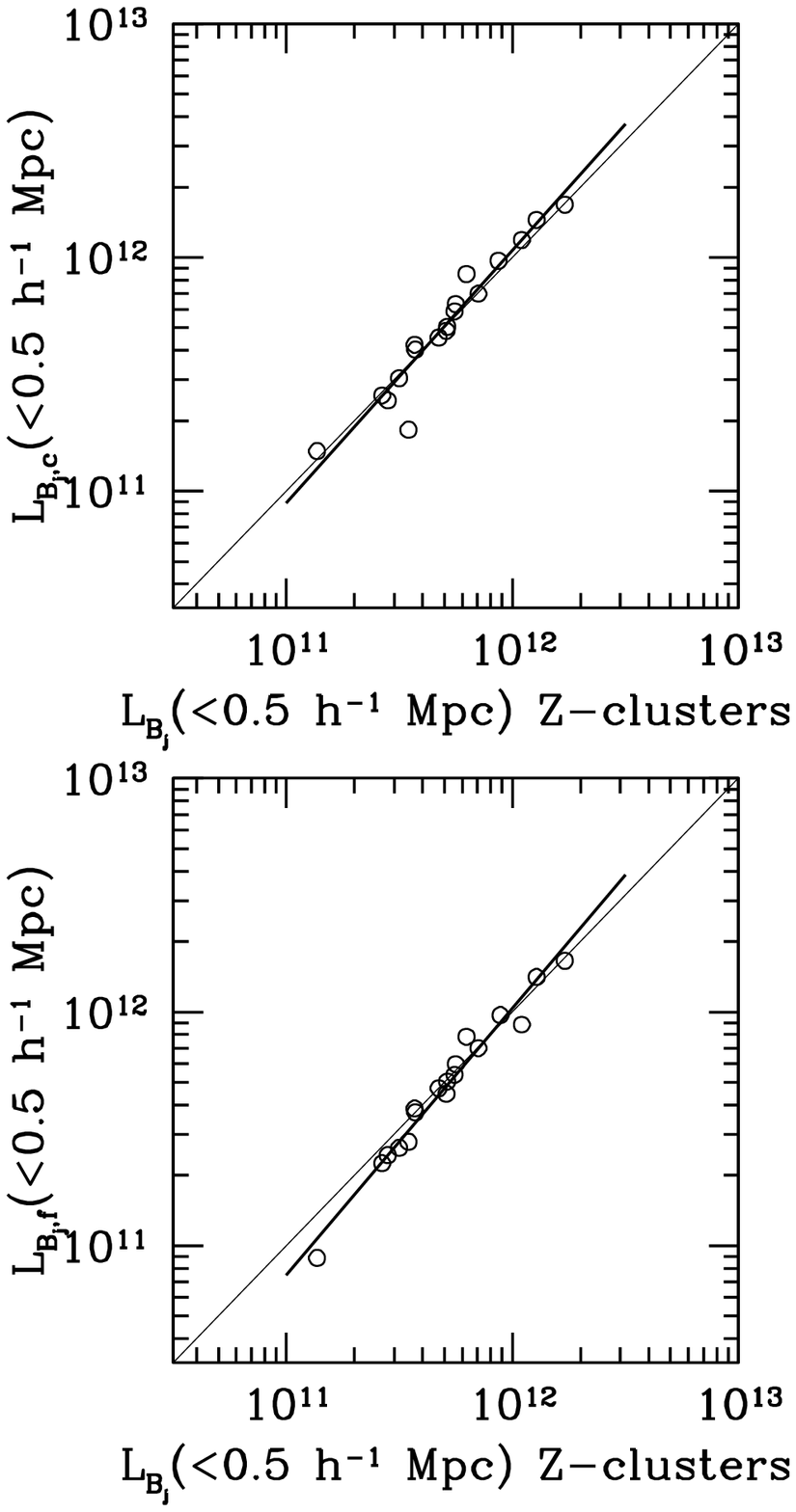}
$\ \ \ \ \ \ $\\
\vspace{9.4truecm}
$\ \ \ $\\
{\small\parindent=3.5mm {Fig.}~2.---
The comparison between luminosities computed for
$Z$--clusters (which do not need fore/background correction) and the
corresponding clusters in $C$-- and $M$-- samples, which have the same
original photometry and which require fore/background correction.  We
consider both $L_{B_j,c}$ (top--panel) and $L_{B_j,f}$ (bottom--panel).  
Luminosities are in units of $h^{-2}~L_{\odot}$.
Heavy
lines represent the linear fits.
}

\vspace{5mm}

As a further test of the robustness of our results, we consider those
51 $C$--clusters whose corresponding redshift samples are based on
ENACS data and compare our derived $f_L$ with those by the original
ENACS analysis of Adami et al. (1998a). They analyzed well sampled
ENACS clusters and found that their luminosity fractions,
$f_L(ENACS)$, are not significantly biased with respect to those
appropriate for COSMOS data.  They found $f_L(ENACS)=0.83$ within a
typical area of 1 \hh$^2$, thus quite similar to our median $f_L$ of
$0.88$ within 0.5 \hh.

As for the estimates of $L_{B_j,c}$, in order to ulteriorly check the
effect of systematic uncertainties from fore/background corrections,
we resort to the background galaxy counts by L97. These counts have
been estimated from the large area ($\sim 0.5$ sr) of the EDSGC, which
is the well--calibrated part of the COSMOS catalogue. L97 found a
general consistency with the previous results by Maddox et
al. (1990b), which is the only other analysis based on a significantly
large area (i.e., the APM sample), where automated procedures to
estimate magnitudes have been applied. A $20\%$ larger estimate is
reported by Colless (1989) who, however, used data from smaller
areas. Even using this larger background estimate, we find that
$L_{B_j,c}$ only decreases by $\lesssim 10\%$.

\subsection{The Extrapolation to Faint Galaxies}

The precise amount of the luminosity correction due to the
extrapolation to faint galaxies depends on the assumed cluster galaxy
LF. Here we assume the usual Schechter form (1976) with parameters
estimated by L97.  As for the LF parameters, there is a general good
agreement among several determination of $M^*$, while significantly
different determinations of the parameter $\alpha$ have been found by
other authors (cf. Rauzy et al. 1998).  Such determinations typically
vary in the range $-1.5,-1$ (see also Marinoni et al. 1999 for a
discussion on this point).  We find for $C$--clusters that the medians
of the luminosity variations are $+18\%$ ($-6\%$) when considering
$\alpha=-1.5$ ($\alpha=-1.0$) instead of $\alpha=-1.22$, as adopted by
L97.

Furthermore, we also consider the possibility of a steepening of the
LF at its faint magnitudes (e.g. Biviano et al. 1995; De Propris \&
Pritchet 1998; Trentham 1998) with respect to a Schechter LF. In
particular, the comprehensive work by Trentham (1998) shows that the
local slope of the LF steepens from $\alpha\sim -1.4$ at $M_B\sim -16$
and $-15$ mag to $\alpha\sim -1.8$ at $M_B\sim -11$ mag. Following
Zucca et al. (1997), we consider here a simplified model based on a
Schechter function and a power law for the high-- and low--luminosity
part of the LF and use their same parameters, $\beta=-1.6$ for the
exponent of the power law and $M_{B_j}=-17$ for the luminosity which
separates the two LF regimes. As a result, we find that luminosities
increases by less than $10\%$.

Our LF extrapolation to faint magnitudes implicitely assumes that
there is no spatial variations of the LF within each cluster and among
different clusters. These assumptions are supported by several
evidences (e.g., Lugger 1986; Colless 1989; Metcalfe, Godwin, \& Peach
1994; Valotto et al. 1997; Trentham 1998; Rauzy et al. 1998; see,
however, Driver, Couch \& Phillips 1998).  In the present analysis we
assume that possible differences lead to random errors which are
comparable to the above $\sim 10\%$ systematics. This point will be
further discussed in \S~6.3 below, when we will discuss possible
systematics affecting the $M$--$L_{B_j}$ relation.

\subsection{Comparison between $C$-- vs. $M$--Luminosities}

We estimate the errors connected with the original photometric samples
(magnitudes and nominal completeness) and with our corrections/conversions 
of magnitudes by comparing luminosities computed for $C$--clusters to
those computed for the same clusters in the $M$--sample.  Figure~3
shows the result of the comparison for the 22 clusters in
common.  There is no evidence of a significant systematic effect,
although within a rather large scatter ($\sim 20\%$).

\subsection{Error Estimates}

From the above analysis we estimate that both random and systematic
errors in the cluster luminosities are $\sim 20-30\%$. Finally, we stress
that our luminosities have been computed by using the cluster centers
reported in Table~2.  After recomputing luminosities for 34
$C$--clusters with X--ray determined centers (Ebeling et al. 1997),
which differ by $\sim 0.1$ \h from the optical centers we adopt in
this work, we find a typical difference of $\sim 5\%$ for luminosities
within $R_{vir}$ and of $\sim 10\%$ for luminosities within 0.5 \hh.
Since the centers we use for luminosities are the same used by G98 for
mass determinations, this kind of error is not taken into account in
our following analysis.
\vspace{-40mm}
\includegraphics{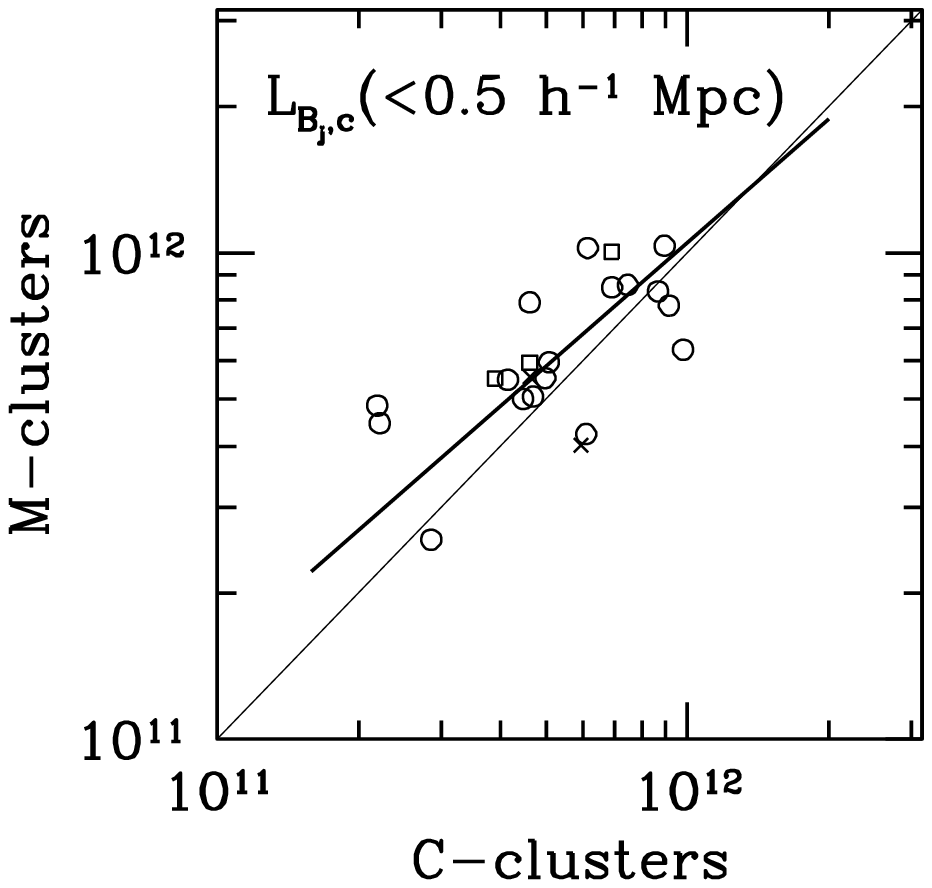}
$\ \ \ \ \ \ $\\
\vspace{9.4truecm}
$\ \ \ $\\
{\small\parindent=3.5mm {Fig.}~3.---
The comparison between $L_{B_j,c}$ luminosities 
computed for $C$--clusters and the corresponding clusters in $M$--
sample, which differ for the original photometry.  
Circles, crosses, and squares denote original blue, visual, and
red  magnitudes, respectively.
Luminosities are units of $h^{-2}~L_{\odot}$.
The heavy line represents the linear fit.
}

\vspace{5mm}

\section{COMPUTING THE MASS--TO--LIGHT RATIOS}

G98 give the virial masses $M$ within $R_{vir}$, and Girardi et
al. (1998a) use these masses to obtain those within 1.5 \h~by
rescaling according to the galaxy distribution, which is assumed to
follow the mass distribution.  Here we obtain masses within 0.5 and
1.0 \h~by following a similar procedure.

By averaging the $M/L_{B_j}$ values for $C$--clusters, we find no
appreciable variation with increasing radius, as expected from the
underlying assumption that mass follows galaxy distribution and from
the scarce relevance of luminosity segregation (e.g. Biviano et
al. 1992; Stein 1997).  For each cluster we report in Table~3 (Col.~2)
the minimum and maximum $M/L_{B_j}$ values among those obtained by
varying the radius, the data sample and the method to estimate
luminosities.  We also give the values of $M/L_{B_j,c}$ and
$M/L_{B_j,f}$ within $R_{vir}$ for clusters belonging to the
$C$--sample (Cols.~3 and~4).

\section{THE RELATIONS BETWEEN LUMINOSITY AND DYNAMICAL QUANTITIES} 

We draw our following results from the analysis of the 89 clusters of
the $C$--sample, which is characterized by a homogeneous photometry
and well defined completeness criteria.  In particular, here we limit
our analysis to mass and luminosities as computed within $R_{vir}$ so
that the correlations between cluster properties will be obtained
at a fixed physical scale.

\subsection{The $L_{B_j}$--$\sigma_v$ relation}

Considering the relation between luminosity and (line--of--sight)
velocity dispersion, $\sigma_v$, has the advantage that 
the latter
quantity is the directly observable one. In fact, the step from
$\sigma_v$ to virial mass estimate requires the additional assumption
that mass distribution follows the galaxy distribution (cf. G98).  We
use the $\sigma_v$ values, which can be considered representative of
the total kinetic energy of galaxies, as given by G98 (see also Fadda
et al. 1996).

\includegraphics{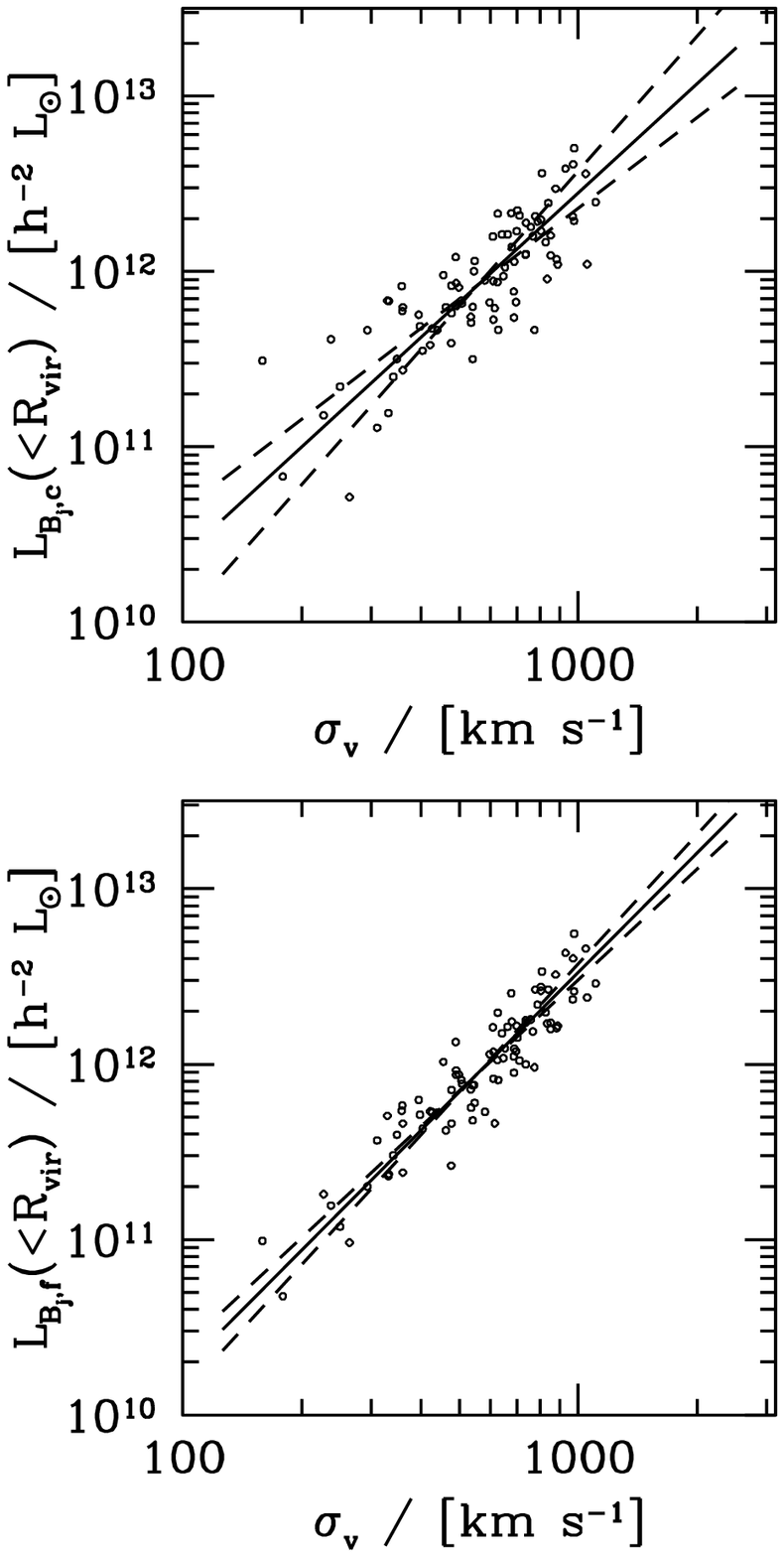}
$\ \ \ \ \ \ $\\
\vspace{9.4truecm}
$\ \ \ $\\
{\small\parindent=3.5mm {Fig.}~4.---
The relation
between luminosity within $R_{vir}$ and velocity dispersion. 
We consider both $L_{B_j,c}$ (top panel) and $L_{B_j,f}$ (bottom panel).
Lines represent the linear fits: dashed lines give the
direct and the inverse
fits, while  the solid line gives the bisecting line.
}

\vspace{5mm}

We fit a regression line in the logarithmic plane,
$L_{B_j}/L_{B_j,\odot}=10^a\sigma^b$ (see Figure~4).
  Using unweighted fits (Isobe et
al. 1990), we calculate both the direct and the inverse, as well as
the bisecting line. Since (random) errors on the variables are
comparable ($\sim 20$--$30\%$ for luminosities and $12\%$ for velocity
dispersions, cf. \S~4.4 and $\ $ G98)  
the $\ $ bisecting line $\ $ should $\ $ provide the 
most appropriate $\ $ regression $\ $ relation. $\ $ We $\ $
obtain $\ $ $a =6.24\pm0.46$ $\ $ and

\end{multicols}

\vspace{6mm} 
\hspace{-4mm}
\begin{minipage}{9cm}
\renewcommand{\arraystretch}{1.2}
\renewcommand{\tabcolsep}{1.2mm}
\begin{center}
\vspace{-3mm} 
TABLE 3\\
\vspace{2mm}
{\sc Mass--to--Luminosity Ratios\\}
\footnotesize
\vspace{2mm} 
\input{tab3a}
\end{center}
\end{minipage}

\vspace{6mm} 
\hspace{-4mm}
\begin{minipage}{9cm}
\renewcommand{\arraystretch}{1.2}
\renewcommand{\tabcolsep}{1.2mm}
\begin{center}
\vspace{-3mm} 
TABLE 3\\
\vspace{2mm}
{\sc continued\\}
\footnotesize
\vspace{2mm} 
\input{tab3b}
\end{center}
\vspace{3mm}  
\end{minipage}

\begin{multicols}{2}

\noindent $b=2.07\pm0.16$ 
for $L_{B_j,c}$ (with $b=1.72$ and $2.56$ for the
direct and inverse fits) and $a =5.73\pm0.27$ and $b =2.26\pm0.10$ for
$L_{B_j,f}$ (with $b=2.10$ and $2.45$ for the direct and inverse
fits).  The scatter is $\sim 25\%$ for $L_{B_j,c}$ and $\sim 15\%$ for
$L_{B_j,f}$.

Alternatively, we also apply a weighted regression line (e.g. Press et
al. 1992), by varying the errors on $L_{B_j,f}$ in the range of
$20$--$30\%$ and considering both a 12\% error in $\sigma_v$ 
and the nominal $\sigma_v$ errors given by G98. We find that $b$
varies in the ranges $2.0$--$2.5$ and $2.2$--$2.4$ for $L_{B_j,c}$ and
$L_{B_j,f}$, respectively.  Moreover, the small scatter in the
$L_{B_j,f}$--$\sigma_v$ relation can be accounted for by the assumed errors.
On the contrary, the large scatter of $L_{B_j,c}$--$\sigma_v$
 can not be
completely accounted for by errors in $L_{B_j,c}$ and $\sigma_v$.

The better defined correlation with $\sigma_v$ suggests that the
$L_{B_j,f}$ values are less affected by random errors than
$L_{B_j,c}$. This is consistent with the fact that $L_{B_j,f}$ is
computed by using local field subtractions, while mean field
corrections are applied for the $L_{B_j,c}$ computation.

\subsection{The $M/L_{B_j}$ ratios and the $M$ -- $L_{B_j}$ relation}

We find $M/L_{B_j,c}=251^{+28}_{-29}$ and
$M/L_{B_j,f}=229^{+22}_{-29}$ for the median values of the
mass-to-light ratios, with errors corresponding to $90\%$
c.l. intervals.

\includegraphics{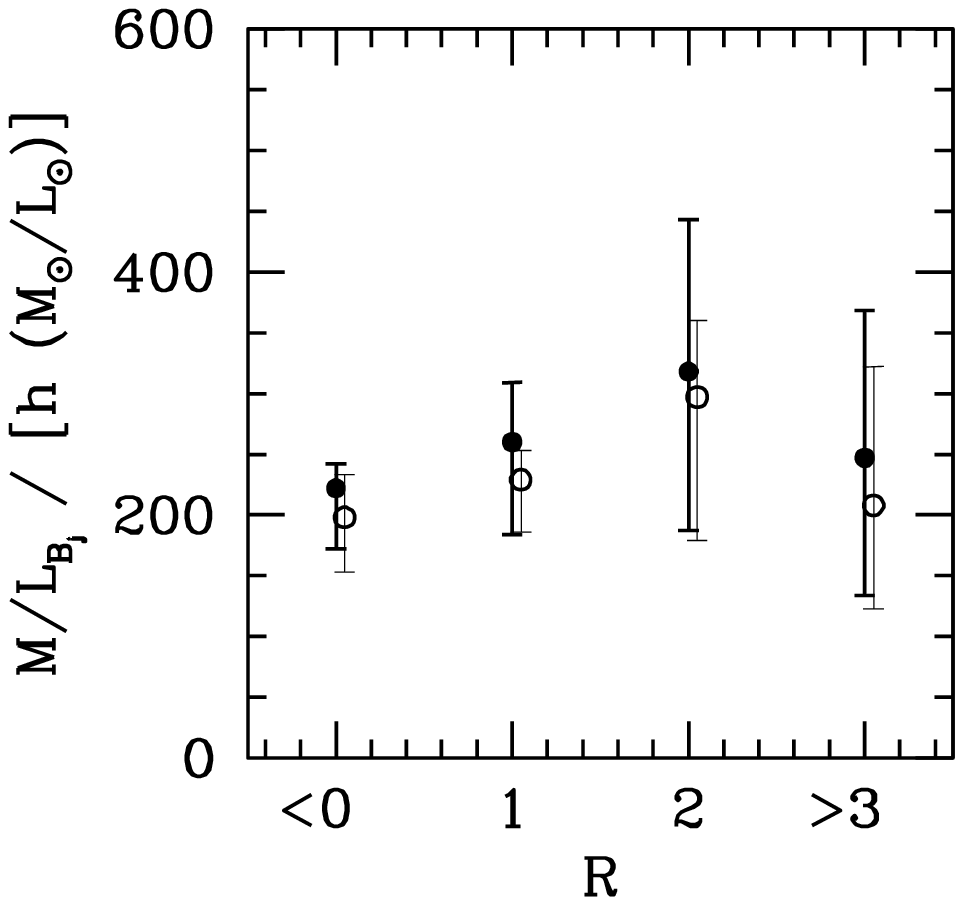}
$\ \ \ \ \ \ $\\
\vspace{5truecm}
$\ \ \ $\\
{\small\parindent=3.5mm {Fig.}~5.---
The behaviour of mass--to--light ratio
vs. cluster richness. Points are median values with $90\%$ c.l. 
error bars. Solid and open points come from $L_{B_j,c}$ and $L_{B_j,f}$ estimates,
respectively.
}

\vspace{5mm}

In order to explore the behavior of $M/L_{B_j}$ with respect to
cluster properties, we look for possible correlations with cluster
morphology, i.e Bautz--Morgan type (86 clusters, types taken from
Abell, Corwin, \& Olowin 1989), Rood--Sastry type (40 clusters, types
taken from Struble \& Rood 1987 and from Struble \& Ftaclas 1994), and
with cluster richness class $R$.  We find only a weak significant
correlation with $R$ (i.e.  $\sim 95\%$ c.l.). Figure~5 shows the
median $M/L_{B_j}$ vs. $R$ where $M/L_{B_j}$ increases with $R$ and
then flattens for the richest $R\ge3$ clusters.

In order to better examine a possible increasing of $M/L_{B_j}$ from
poor to rich structures, we avoid analyzing the behavior of
$M/L_{B_j}$ vs. $M$ or $L_{B_j}$ because $M/L_{B_j}$ is defined as a
function of $M$ and $L_{B_j}$, and therefore we were plotting
correlated quantities (see, e.g.,  Mezzetti, Giuricin, \&
Mardirossian 1982). Instead, we directly examine the $M$--$L_{B_j}$
relation by fitting a regression line in the logarithmic plane,
\begin{equation}
{M\over M_{\odot}}\,=\,10^c\cdot \left({L_{B_j}\over L_{B_j,\odot}}\right)^d\,.
\label{eq:ml}
\end{equation}

As for the $L_{B_j,f}$ luminosity, we obtain $c =-0.34\pm0.72$ and $d
=1.22\pm0.06$ for the unweighted bisecting fit, with a scatter of
$\sim 30\%$ (see Figure~6).  As for the weighted regression fit, by
varying the errors on $L_{B_j,f}$ in the range $20$--$30\%$ and considering
a global error on $M$ by $40\%$, or alternatively the nominal mass
errors given by G98, we obtain that $d$ varies in the range
$1.17$--$1.23$, again significantly larger than unity.

\includegraphics{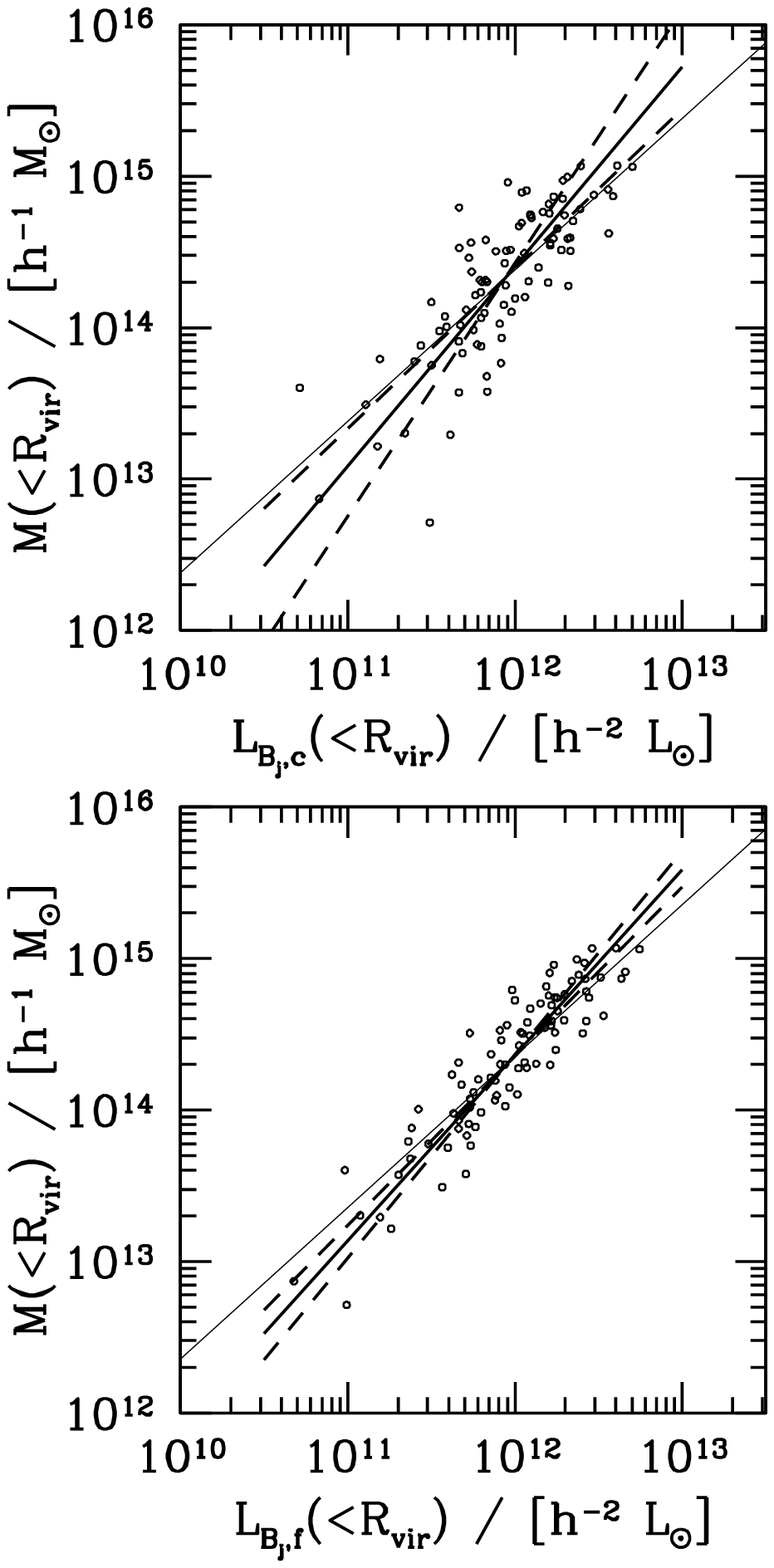}
$\ \ \ \ \ \ $\\
\vspace{9.4truecm}
$\ \ \ $\\
{\small\parindent=3.5mm {Fig.}~6.---
The relation
between mass and luminosity within $R_{vir}$. 
We consider both $L_{B_j,c}$ (top panel) and $L_{B_j,f}$ (bottom panel).
Heavy lines represent the linear fits: dashed lines give the
direct and the inverse
fits, while  the solid line gives the bisecting line.
The faint line is the $M\propto L_{B_j}$ fitted relation.
}

\vspace{5mm}

As for the $L_{B_j,c}$ luminosity, the fitted bisecting line gives $c
=-1.41\pm1.30$ and $d =1.32\pm0.11$ with a scatter of
$\sim45\%$. The weighted analysis gives $d=1.06$--$1.33$ and,
again, the large scatter in the $M$--$L_{B_j,c}$ relation is not completely
justified by the assumed errors.

\subsection{The Robustness of the $M$--$L_{B_j}$ Relation}
We have discussed in \S~4 the possible systematics affecting the
estimates of cluster luminosities. Although we have shown that such
systematics should not introduce strong biases in $L_{B_j}$, it
is anyway worth investigating their effects on the $M$--$L_{B_j}$
relation, being the tendency of cluster mass to increase faster than
luminosity a small (although significant) effect
[cf. eq.(\ref{eq:ml})].

As for the background corrections, we first note that, although the
methods for their estimate for $L_{B_{j,c}}$ and $L_{B_{j,f}}$ are
completely independent, the two resulting mass-luminosity relations
($M$--$L_{B_j,c}$ and $M$--$L_{B_j,f}$) well agree both in slope and
in normalization.

An increase in the amount of mean field counts for the $L_{B_j,c}$
correction has the effect of reducing more the luminosities of poorer
clusters. For instance, the use of the 20\% larger background
correction by Colless (1989) decreses the exponent $d$ in
eq.(\ref{eq:ml}) only from 1.32 to 1.28. In order to make it
consistent with unity at 2-sigma c.l., one has to assume that the L97
counts underestimate the true background by about $40\%$. This seems
rather unlikely, in view of the similarity between the results by L97
and Maddox et al. (1990b), which are the two analysis based on the
largest analyzed area.

A possible way to change the slope of the $M$--$L_{B_j,f}$ relation
would be to have a dependence on the cluster mass for the depth of the
redshift samples, from which the $f_L$ fractions are estimated.
However, when using the homogeneous subsample of 51 clusters for which
the $f_L$ values are obtained from ENACS data (cf. \S~4.1) we find the
same $M$--$L_{B_j,f}$ relation, with slope $d=1.21$.

A larger increase of luminosities for richer clusters can also be the
consequence of more negative $\alpha$ values in the Schechter LF which
is used to extrapolate  the cluster luminosity to faint galaxy
magnitudes (cf.  \S~4.2). For $\alpha=-1.5$ we obtain $d=1.20$ and
$1.28$ for $L_{B_j,f}$ and $L_{B_j,c}$, respectively, while
$\alpha=-1.9$ is required to reconcile $d$ with unity within 2-sigma
c.l. However, current observations indicates that the LF can be so
steep only at very faint magnitues.  When using the LF model discussed in
\S~4.2, i.e. a Schechter function coupled with a power--law, we still
find $d>1$ at 2.5--3 sigma c.l.  even considering the rather extreme
case $\alpha=-1.5$ and $\beta=-2$ for the exponents of the Schechter
and power--law parts of the LF, respectively.

As for the effect of possible systematic variations of the LF both
from cluster to cluster and within each cluster, several works
considered the dependence of the galaxy LF on cluster radius and found
no significant changes (e.g. Colless 1989; Metcalfe et al.
1994; Valotto et al. 1997; Rauzy et al. 1998), thus in agreement with
evidences that luminosity segregation concerns only very luminous
galaxies (e.g. Biviano et al. 1992; Stein 1997). A recent analysis of
seven clusters at $z\sim 0.15$ (Driver et al.  1998) shows that the
fraction of dwarf--to--giant galaxies depends on the physical region
analyzed and increases significantly from the inner to the outer
cluster region. However, we expect this not to represent a serious
concern for our analysis, since our results are based on luminosities
all computed within the physical radius $R_{vir}$.

A relevant point in the determination of the slope of the
$M$--$L_{B_j}$ relation concerns the universality of the
LF. Indeed, a flatter LF for less massive clusters could in principle
reconcile the observed relation with $M\propto L_{B_j}$. L97 found no
clear evidence for LF variations with the Bautz--Morgan class, $BM$,
or richness (see also Lugger 1986; Colless 1989; Trentham 1998; and
recently Rauzy et al. 1998 for ENACS clusters).  However, there is a
tendency to find that clusters with a dominant cD galaxy and/or
$BM=1$, high richness and regular $X$--ray emission have flat LF
(e.g. Dressler 1978b; L\'opez et al. 1997; see however Valotto
et al. 1997). From a recent analysis of seven clusters down to faint
magnitudes ($M_R=-16.5$), Driver et al.  (1998) found that the ratio
of dwarf--to--giant galaxies decreases (from 3.1 to 0.8) with the mean
local projected galaxy density (cf. also Garilli, Maccagni, \& Andreon
1998). This turns into a flatter LF in more dense environments, with
$\alpha $ varying from --0.9 to --1.4.  In general, the global cluster
parameter which correlates with the LF slope seems to be the
morphological type, rather than the cluster richness (e.g., Driver et
al. 1998). Indeed, there is a well known correspondence among
different cluster classifications, where regular clusters are usually
of early Bautz-Morgan or Rood-Sastry types, with high central galaxy
density and elliptical--rich systems (see Bahcall 1977 for a
review).

A clear evidence for a correlation between cluster regularity (or
density) and mass has been not found to date, although indications in
this direction have been provided, for instance, by the correlation
between regularity and $X$--ray luminosity (e.g. Forman \& Jones 1982
for a review). Therefore, if any, the effect of a correlation between
cluster mass and LF slope would go in the direction of increasing the
slope $d$ of the $M$--$L_{B_j}$ relation.

Finally, we consider the possibility that cluster substructures can
contribute to increase the scatter or to change the slope of the
$M$--$L_{B_j}$ relation.  We extract only those 54 clusters for
which the significance for the presence of substructures is $<90\%$,
according to a Dressler--Schectman test (Dressler \& Schectman 1988)
applied to the reference redshift samples of G98. We find consistent
results for the slope of the $M$--$L_{B_j}$ relation and no reduction
of its scatter.

\section{DISCUSSION}

\subsection{The $L$--$\sigma_v$ Relation}

Our result that $L_{B_j} \propto \sigma_v^{\mbox{\rm 2.1--2.3}}$, for
$L_{B_j}$ estimated within the virialization radius $R_{vir}$, can be
compared to previous determinations, at
least for those cases where luminosities are measured within some
physical radius.  Schaeffer et al. (1993), who used the total cluster
luminosities by West, Oemler, \& Dekel (1989), found $L_V \propto
\sigma_v^{1.87\pm 0.44}$.  Adami et al. (1998a), who estimated
luminosities within a square of 10 core--radii of diameter, found
$L_{B_j} \propto \sigma_v^{\mbox{\rm 1.56}}$ but with a large scatter.
Steeper relations are found for a variety of smaller systems, e.g.
$L_V\propto \sigma_v^{2.3}$ for globular clusters (Paturel \& Garnier
1992), $L_B\propto \sigma_v^{2.5}$ for dwarf ellipticals (Held et
al. 1992), and $L\propto \sigma_v^ {\mbox{\rm 3.5--4}}$ for
ellipticals (e.g. Faber \& Jackson 1976; Dressler et al. 1987).

Some authors, by analyzing clusters in analogy with elliptical
galaxies, found a relation betwen luminosity, cluster size $R$ and
velocity dispersion of the type $L\propto R^A \sigma_v^B$.  Typical
values for the parameters defining this cluster fundamental plane
range in the intervals $A=0.9$--1.4 and $B=0.9$--1.3 (Schaeffer et
al. 1993; Adami et al. 1998a).  Since for a virialized structure we
expect its scale to be proportional to the internal velocity
dispersion, the above scaling relations would traslate into $L \propto
\sigma_v^{\mbox{\rm 1.8--2.7}}$ thus in agreement with our findings.
A detailed fundamental plane analysis for our cluster sample will be
deserved to a forthcoming paper.

\subsection{The $M$--$L$ Relation}

One of the main results of our analysis is that mass has a slight, but
significant, tendency to increase faster than the luminosity.  In
particular, we find that $M\propto L_{B_j}^{\mbox{\rm 1.2--1.3}}$,
which translates into a $M/L_{B_j}$ variation of a factor of $2$--$3$
from poor to rich clusters. Again, this result can be related to
those inferred from the analysis of the cluster fundamental plane. In
fact, assuming the virialization state and internal structure of all
clusters to be identical, Schaeffer et al. (1993) found that
$M/L_V\propto L_V^{0.3}$ and Adami et al. (1998a) showed that
$M/L_{B_j}\propto \sigma_v$ (i.e. $\propto M^{1/3}$ for the mass
contained within the virialized region, cf. G98). In general there is
no evidence of correlation between $M/L$ and cluster properties
(e.g. Dressler 1978b; Carlberg et al. 1996; Fritsh \& Buchert
1999; see, however, Adami et al. 1998b, who found a correlation with
$\sigma_v$). Due to the small size of the effect we detected, it is
clear that it can be seen only in a rather large cluster sample spanning
a large dynamical range.

In \S~6.3 we verify the robustness of the $M$--$L_{B_j}$ against
possible systematics. Here we analyze the possibily that the fitted
slope might be completely driven by the kind of photometry of our
sample.  Since late--type galaxies are bluer than early--type
galaxies, a possible over-abundance of late--type galaxies in poor
clusters could flatten the observed relation between masses and
luminosities coming from $V$ or $R$ photometry.

It is well known that the late--type galaxy fraction decreases with
the local density (Dressler et al. 1980) and increases with the
distance from the cluster center (Whitmore, Gilmore, \& Jones
1993). Since, as already discussed in \S~6.3, there is no well
established correlation between cluster density and mass, it is
difficult to establish a morphology--richness correlation. In fact,
Whitmore et al. (1993) found that the behaviour of morphology vs.
radius is independent of $\sigma_v$.  However, if we assume that this
relation does indeed exist and poor clusters are really
spiral--richer, we can attempt a simple calculation of the effect of
working with $R$ photometry instead with $B$ one (since we use $B_j$
instead of $B$ the true effect would be also smaller). Since
$B_{\odot}=5.48$ and $R_{\odot}=4.31$ (Allen 1973) and being
$(B-R)\sim 1.5$ for early type galaxies and $(B-R)\sim 1$ for spirals
of de Vaucouleurs type 5 (Buta et al. 1994 and Buta \& Williams 1995),
we find that $L_R^e=1.3 L_B^e$ for luminosity of early--type galaxies
and $L_R^l=0.8 L_B^l$ for luminosity of late type galaxies.  Then,
when assuming that typical galaxy luminosities in clusters are roughly
comparable for early and late type galaxies (e.g. Sandage, Binggeli,
\& Tammann 1985; Andreon 1998) and that the ratio between
early-to-late type galaxies goes from 3.5 to 1 for spiral--poor and
spiral--rich clusters, respectively (e.g. Oemler 1974), we obtain
$L_R\sim 1.20 L_B$ and $L_R\sim 1.05 L_B$ for spiral--poor and
spiral--rich clusters, respectively. Even assuming that these
different conversions must be applied to clusters whose mass is larger
and smaller that the median value, respectively, we obtain $M\propto
L_R^d$ with $d>1$ at 2.5--sigma c.l.

Therefore, although we suggest that a conclusive result could be found
only by directly analyzing other samples in different magnitude bands,
we find at present no evidence that our mass--dependent $M/L$ relation
can be explained by a higher spiral fraction in poorer clusters.

\subsection{The $M/L$ ratios}
Median values of $M/L_{B_j}$ are found to range in the interval
230--250 \ml, thus in good
agreement with recent estimates (e.g. David et al. 1995)
and definitely lower than older estimates (cf.  Faber
\& Gallagher 1979).  Our $M/L_{B_j}$ ratios are smaller
than those by Adami et al. (1998b).  Since their luminosities are
comparable to ours, this difference is due to the smaller masses fund
by G98.  Our $M/L_{B_j}$ ratio agrees with that by Carlberg et
al. (1996, $M/L_R\sim 300$), whose optical virial masses are have been
shown to be consistent with X--ray masses (Lewis et al. 1999; note
that the change of luminosity due to evolutionary effects roughly
cancels the change of luminosity betweeen $R$ and $B$ bands).

The assumption that $M/L$ within clusters is typical of the Universe
as a whole leads to an estimate of the matter density parameter
$\Omega_m$ (hereafter $(\Omega_m)_{M/L})$, e.g. Bahcall et al. 1995;
Carlberg et al. 1996), which has the advantage of being independent of
the value of $H_0$ and of the presence of a cosmological constant.
Estimates of the $B$--band luminosity density gives $\rho_L\simeq
2\times 10^8h\,L_\odot$ Mpc$^{-3}$ (Efstathiou, Ellis \& Peterson
1988; Zucca et al. 1997; Loveday, Tresse \& Maddox 1999), which
corresponds to the $M/L_B$ closure value of $\simeq 1350$
\ml. Although we take this value in the following as the reference
one, we remind that its determination is quite sensitive to
uncertainties in the local galaxy luminosity function (cf.  Marinoni
et al. 1999). With the above value for the closure $M/L_B$, we obtain
$(\Omega_m)_{M/L}\simeq 0.2$, in agreement with recent determinations
of $\Omega_m$ based, for instance, the $M/L$ ratio of distant CNOC
clusters (Carlberg et al. 1996), on the gas fraction within clusters
(e.g. Evrard 1997) and on combining high--$z$ SN$Ia$ data with
preliminary measurements of small scale CMB anisotropies (e.g. Tegmark
1999).

As already noted by Rees (1985), an increase of $M/L$
with the scale of the collapsed structure within which it is estimated
can in principle reconcile the observed cluster $M/L$ values
with $\Omega_m=1$. This could be possible in the
framework of biased galaxy formation (e.g. Kaiser 1984; Bardeen et al
1986) and, indeed, various physical mechanisms could produce
large--scale variations in the conversion efficiency from baryons to
luminous galaxies (e.g. Rees 1985; Blumenthal et al. 1984;
Dekel \& Rees 1987) leading to $(\Omega_m)_{M/L}\le
\Omega_m$. Although the difference between efficiencies of
gas conversion into stars for field and clusters is not yet precisely
known, reconcilying such low $M/L$ values with a critical density
Universe would require a difference in such efficiencies by at least a
factor three. 
Furthermore, the mild increase of $M/L$ with cluster luminosity that   
we find is more consistent with a low--density and low--bias model, 
rather than with a high--bias critical--density model.

\section{SUMMARY AND CONCLUSIONS}

We analyze a sample of 105 galaxy clusters for which Girardi et
al. (1998b, G98) computed
virial masses, all having available galaxy magnitudes with some degree
of completeness. In particular, our main result on the cluster
mass--to--light ratio are drawn from a subsample of 89 clusters with
$B_j$ band galaxy magnitudes taken from the survey of COSMOS/UKST
Southern Sky Object Catalogue (Yentis et al. 1992). Photometry for
other clusters is provided at different magnitude bands.

After suitable magnitude corrections and uniform conversions to the
$B_j$ band, we compute cluster luminosities within several
clustercentric radii, $0.5,1.0,1.5$ \hh, and within the virialization
radius $R_{vir}$. To this purpose, we use the results by Lumsden et
al. (1997, L97) on the luminosity functions and galaxy counts for the
EDSGC, which is the well-calibrated part of the COSMOS catalogue.

We analyze the effect of several uncertainties connected with original
photometric data, our corrections/conversions of galaxy magnitudes,
fore/background subtraction, and extrapolation below the completeness
limit of the photometry. In particular, we find an overall agreement
between luminosities computed by applying two independent methods of
fore/background correction, one based on mean galaxy counts,
$L_{B_j,c}$, and one taking into accounts the local cluster field,
$L_{B_j,f}$. We find that both random and possible systematic errors
on luminosities are about $20-30\%$.  We compute and list $M/L$ ratios
for the 105 clusters.

In order to examine the relations between cluster luminosities and
dynamical quantities we consider only the homogeneus sample
($C$--sample) taken from the COSMOS catalogue. In this way, we avoid
possible errors connected to magnitude band conversions and have the
advantage of dealing with a well defined magnitude completeness. Using
the COSMOS magnitudes is also consistent with our choice to use the
L97 results on luminosity function and galaxy counts in the luminosity
calculations.  Cluster luminosities are computed within the
virialization radius for all clusters.

Our main results can be summarized as follows.

(a)~We find a well--defined correlation between cluster luminosity and
galaxy velocity dispersion, $L\propto \sigma_v^{\mbox{\rm 2.1--2.3}}$,
with a scatter of $\sim 15\%$ and $\sim 25\%$ for $L_{B_j,f}$ and
$L_{B_j,c}$, respectively. As expected the luminosities $L_{B_j,f}$
based on a local field correction shows smaller random errors than
$L_{B_j,c}$, which is based on a mean field correction.

(b)~Different methods to estimate luminosities provide median
mass--to--light ratios with typical values $M/L_{B_j} \sim 230$--$250$ \ml.
  We do not find any correlation of $M/L_{B_j}$ with cluster
morphologies, i.e. Rood--Sastry and Bautz--Morgan types, and only a
weak significant correlation with cluster richness.

(c)~We find a good correlation between cluster luminosity and mass.
In particular, mass has a slight, but significant, tendency to
increase faster than the luminosity does, 
$M\propto L_{B_j}^{\mbox{\rm 1.2--1.3}}$. 
 The scatter is $\sim 30\%$ and $\sim 45\%$ for
$L_{B_j,f}$ and $L_{B_j,c}$, respectively.  We verify the robustness
of this relation and show that the $M \propto L_{B_j}$ relation is
very difficult to be reconciled with available data.

Finally, by using a simple approach, we verify that the tendency of
mass to increase faster than the luminosity cannot be explained by a
higher spiral fraction in poorer clusters. Therefore, we suggest that
a similar result should also be found when analyzing a cluster sample
with $R$ band galaxy magnitudes. However, a more conclusive result on
this point should wait for a direct analysis of $R$ band magnitude
samples.

\acknowledgments 
We thank F. Durret and N. Metcalfe for having given
us the electronic version of their catalogs.  We are particularly in
debt with Sandro Bardelli for having giving us his compilation of
A3558 cluster and for enlighting suggestions, too.  We also thank
Andrea Biviano, Peter Katgert, Massimo Persic, Massimo Ramella, Paolo Salucci,
and Dario Trevese, for useful
discussions. Special thanks to Alessandro Cristofoli for his help in
initial phase of this project.  
We thank an anonymous referee for useful suggestions.
S.B. wishes to acknowledge
Osservatorio Astronomico and ICTP in Trieste for the hospitality during the
preparation of this work.  This research has made large use of the
COSMOS/UKST Southern Sky Catalogue supplied by the Anglo--Australian
Observatory.  This work has been partially supported by the Italian
Ministry of University, Scientific Technological Research (MURST), and
by the Italian Space Agency (ASI).

\end{multicols}


\end{document}

%% file: tab1a.tex
%
%
  
\begin{tabular}{lrllr}
\hline \hline
\multicolumn{1}{c}{Name}
&\multicolumn{1}{c}{Refs.}
&\multicolumn{1}{c}{Band}   
&\multicolumn{1}{c}{$m_{lim}$}  
&\multicolumn{1}{c}{$\%$ Compl.}  
\\
\hline
\hline
\multicolumn{5}{c}{$M$--Sample -- Blue Magnitudes}\\   
\hline
A85     & 1 & $B_j$        &  19.75   & 100  \\
A194    & 2 & $B_{Zwicky}$ &   15.6   &  93  \\
A426    & 3 & $B_{Zwicky}$ &   15.7   & 100  \\
A458    & 4 & $B_j(APM)$        &   20.0   & 100  \\
A539    & 5 & $B_{Zwicky}$ &   15.7   &  91  \\
A1656   & 6 & $B$          &   20.0   & 100  \\
A1656   & 7 & $B_{Zwicky}$ &  15.7    & 100  \\
A2197   & 8 & $B_{Zwicky}$ &   15.7   & 100  \\
A2554   & 4 & $B_j(APM)$        &   20.2   & 100  \\
A2670   & 9 & $B_j$        &   19.0   & 100  \\
A2717   & 4 & $B_j(APM)$        &   18.9   & 100  \\
A2721   & 4 & $B_j(APM)$        &   19.7   & 100  \\
A3126   & 4 & $B_j(APM)$        &   18.8   & 100  \\
A3128   & 4 & $B_j(APM)$        &   18.5   & 100  \\
A3334   & 4 & $B_j(APM)$        &   20.0   & 100  \\
A3360   & 4 & $B_j(APM)$        &   20.0   & 100  \\
A3558   &10 & $B$          &   20.85  & 100  \\
A3574   &11 & $B$          &   16.0   & 100  \\
A3667   &12 & $B_j$        &   18.0   & 100  \\
A3705   & 4 & $B_j(APM)$        &   20.1   & 100  \\
S0753   &11 & $B$          &   16.6   & 100  \\
S1157   & 4 & $B_j(APM)$        &   19.2   & 100  \\
AWM1    &13 & $B_{Zwicky}$ &   15.7   & 100  \\
AWM7    &13 & $B{_Zwicky}$ &   15.7   & 100  \\
Virgo   &14 & $B$          &   15.4   & 100  \\
\hline
\multicolumn{5}{c}{$M$--Sample -- Visual Magnitudes}\\   
\hline
A85 & 1 & $V$          &  18.75   & 100  \\
A1060   &15 & $V$          &  16.65   & 100  \\
\hline
\multicolumn{5}{c}{$M$--Sample -- Red Magnitudes}\\	    
\hline
A85 & 1 & $R$          &  18.25   & 100  \\
A119   &16 & $R$          &  17.8    & 100  \\
A539 & 5 & $R$          &  16.83   & 100  \\
A576   &17 & $R$          &  17.0    &  86  \\ 
A1314   &18 & $F$          &  16.1    & 100  \\
A2040   &19 & $F$          &  17.0    & 100  \\
A2151   &20 & $R$          &  15.1    & 100  \\ 
A2256   &21 & $r$          &  16.8    & 100  \\
A2593   &19 & $F$          &  16.9    & 100  \\
A2670   &19 & $F$          &  17.6    & 100  \\
AWM4    &22 & $R$          &  17.5    & 100  \\
MKW4    &22 & $R$          &  16.5    & 100  \\
\hline
\multicolumn{5}{c}{$Z$--Sample}\\			    
\hline
A0194	& 2 & $B_{Zwicky}$ &    15.6  &  93  \\ 
A0426	& 3 & $B_{Zwicky}$ &    15.7  & 100  \\ 
A0576	&17 & $R$          &    17.0  &  86  \\ 
A1060	&15 & $V$          &    14.9  & 100  \\ 
A1656	& 7 & $B_{Zwicky}$ &    15.7  & 100  \\ 
A2151	&20 & $R$          &    15.1  & 100  \\ 
A2197	& 8 & $B_{Zwicky}$  &    15.7  & 100  \\ 
A2256	&21 & $r$          &    16.7  & 100  \\ 
A2670	& 9 & $B_j$        &    19.0  & 100  \\ 
A3558	&23,24 & $B_j(COSMOS)$     &    18.0  &  88  \\ 
\hline
\end{tabular}

%% file: tab1b.tex
%
%
  
\begin{tabular}{lrllr}
\hline \hline
\multicolumn{1}{c}{Name}
&\multicolumn{1}{c}{Refs.}
&\multicolumn{1}{c}{Band}   
&\multicolumn{1}{c}{$m_{lim}$}  
&\multicolumn{1}{c}{$\%$ Compl.}  
\\
\hline
\hline
A3574   &11 & $B$          &    15.5  & 100  \\ 
A3667   &12 & $B_j$        &    17.5  &  91  \\ 
S0753   &11 & $B_j$        &    15.5  & 100  \\ 
AWM1	&13 & $B_{Zwicky}$ &    15.7  & 100  \\ 
AWM4	&   & $R$          &    15.5  & 100  \\ 
AWM7	&13 & $B_{Zwicky}$ &    15.7  & 100  \\ 
MKW4	&   & $R$          &    15.5  &  94  \\ 
Ursa Major &25 & $B$          &    14.45 & 100  \\ 
Virgo	&14 & $B$          &    15.4  &  89  \\          
\hline
\end{tabular}

%% file: comm_tab1.tex
{\footnotesize\parindent=3mm
(1)  Slezak et al. 1998; (2) Chapman et al. 1988 ; (3) Kent \&
Sargent 1983; (4) Colless 1989; (5) Ostriker et al. 1988; 
(6) Godwin, Metcalfe, \& Peach 1983; (7) Kent \& Gunn 1982;
(8) Gregory \& Thompson 1984; (9) Sharples, Ellis, \& Gray 1988;
(10) Metcalfe et al. 1994; (11) Willmer et al. 1991;
(12) Sodr\'e et al. 1992; (13) Beers et al. 1984; 
(14)  Binggeli, Sandage, \& Tammann 1985; (15) Richter 1989;
(16) Fabricant et al. 1993; (17) Mohr et al. 1996; (18) Flin et al. 1995;
(19) Trevese et al. 1997; (20) Barmby \& Huchra 1998; 
(21) Fabricant, Kent, \& Kurtz 1989; (22) Malumuth \& Kriss 1986;
(23) Bardelli et al. 1998;
(24) COSMOS catalog; (25) Tully et al. 1996}

%% file: tab2a.tex
%
%

\begin{tabular}{lcrrcrcrrrrc}
\hline \hline
\multicolumn{1}{c}{Name}
&\multicolumn{1}{c}{Sample}
&\multicolumn{1}{c}{Center}
&\multicolumn{1}{c}{$N(0.5)$ }   
&\multicolumn{1}{c}{$L_{B_j}(0.5)^a$}  
&\multicolumn{1}{c}{$N(1.0)$ }   
&\multicolumn{1}{c}{$L_{B_j}(1.0)^a$}  
&\multicolumn{1}{c}{$N(1.5)$ }   
&\multicolumn{1}{c}{$L_{B_j}(1.5)^a$}  
&\multicolumn{1}{c}{$R_{vir}$ }   
&\multicolumn{1}{c}{$N(R_{vir})$ }   
&\multicolumn{1}{c}{$L_{B_j}(R_{vir})^a$}
\\
\multicolumn{1}{c}{}
&\multicolumn{1}{c}{}
&\multicolumn{1}{c}{$\alpha(1950)-\delta(1950)$}
&\multicolumn{1}{c}{}   
&\multicolumn{1}{c}{$10^{11} h^{-2} L_{\odot}$}  
&\multicolumn{1}{c}{}   
&\multicolumn{1}{c}{$10^{11} h^{-2} L_{\odot}$}  
&\multicolumn{1}{c}{}   
&\multicolumn{1}{c}{$10^{11} h^{-2} L_{\odot}$}  
&\multicolumn{1}{c}{$h^{-1}\ Mpc$}  
&\multicolumn{1}{c}{}   
&\multicolumn{1}{c}{$10^{11} h^{-2} L_{\odot}$}  
\\
\hline
A85               & $C$       &003908.9-093501&   63&    4.6,    5.2&  164&    7.8,    8.6&  287&   15.1& 1.94&  403&   20.5,   23.4\\
A85               & $M$(Blue) &003908.9-093501&  165&    7.9,    8.5&  419&   17.3,   16.5&  741&   28.8& 1.94& 1036&   40.1,   38.3\\
A85               & $M$(Vis)  &003908.9-093501&   94&    5.6,    6.2&  242&   11.5,   11.7&  421&   18.6& 1.94&  581&   26.3,   28.0\\
A85               & $M$(Red)  &003908.9-093501&  103&    5.9,    6.6&  265&   12.3,   12.4&  453&   19.9& 1.94&  628&   28.4,   29.8\\
A119              & $C$       &005344.7-013143&  110&    3.9,    4.5&  263&    9.8,   11.7&  400&   13.8& 1.36&  370&   13.8,   17.5\\
A119              & $M$(Red)  &005344.7-013143&   90&    5.5,    6.0&  211&   11.9,   13.4&  288&   14.0& 1.36&  266&   13.7,   17.1\\
A194              & $C$       &012319.6-013545&  148&    2.2,    2.4&  427&    2.8,    3.9&  931&    2.8&  .68&  218&    2.5,    3.0\\
A194              & $M$(Blue) &012319.6-013545&   25&    4.9,    4.5&   47&    8.6,    7.3&   62&   10.6&  .68&   33&    6.3,    5.9\\
A194              & $Z$       &012319.6-013545&   25&    5.1&   42&    8.4&   50&   10.1&  .68&   33&    6.7\\
A229              & $C$       &013642.7-035317&   25&    4.6,    5.5&   50&    6.9,    8.1&   80&    7.4& 1.01&   50&    6.8,    8.1\\
A256              & $C$       &014533.0-040927&   25&    2.0,    2.3&   75&    4.9,    4.5&  134&   13.4& 1.09&   85&   10.0,    7.6\\
A295              & $C$       &015938.9-012032&   76&    3.1,    2.9&  227&    8.8,    9.0&  466&   17.2&  .72&  137&    5.9,    5.8\\
A420              & $C$       &030653.2-114353&   29&    2.0,    1.6&   72&    3.7,    3.5&  120&    4.3&  .72&   45&    2.7,    2.4\\
A426              & $M$(Blue) &031621.9+412156&   40&   11.9,    8.8&   60&   15.8,   12.7&   79&   17.8& 2.05&  105&   20.3,   19.8\\
A426              & $Z$       &031621.9+412156&   40&   11.0&   59&   16.2&   73&   19.3& 2.05&   94&   24.0\\
A458              & $C$       &034357.0-243001&   27&    5.0,    5.7&   68&    8.3,   10.1&  118&   13.7& 1.47&  114&   12.4,   17.8\\
A458              & $M$(Blue) & 34357.0-243001&   58&    5.5,    6.5&  155&   11.3,   13.5&  -& -& 1.47&  -& -\\
A496              & $C$       & 43112.9-132322&  133&    2.6,    2.8&  319&    5.2,    6.8&  610&   10.6& 1.37&  526&    7.7,   11.1\\
A514              & $C$       & 44620.3-203847&   55&    3.1,    3.7&  146&    8.6,   10.7&  249&   10.9& 1.76&  294&   11.7,   16.1\\
A524              & $C$       & 45532.7-194823&   33&    2.2,    1.2&   71&    2.8,    1.4&  117&    3.2&  .50&   33&    2.2,    1.2\\
A539              & $M$(Blue) & 51355.7-062436&    6&    7.3,    6.4&   13&   13.5,   11.6&   22&   20.0& 1.26&   18&   16.6,   14.4\\
A539              & $M$(Red)  & 51355.7-062436&   53&    3.2,    2.5&   85&    4.6,    3.9&  -& -& 1.26&  -& -\\
A576              & $M$(Red)  & 71719.1+555221&   79&    5.9,    5.4&  180&   11.7,   10.0&  -& -& 1.83&  -& -\\
A576              & $Z$       & 71719.1+555221&   72&    5.5&  141&   10.7&  -& - & 1.83&  -& -\\
A978              & $C$       &101758.0-061546&   57&    2.3,    2.8&  149&    5.3,    6.7&  219&    5.2& 1.07&  161&    5.5,    7.2\\
A1060             & $C$       &103410.1-2714 2&  782&    5.9,    5.8& 2533&    8.7,   10.3& 5310&    9.1& 1.22& 3678&    8.8,   11.8\\
A1060             & $M$(Vis)  &103410.1-271402&  126&    4.0,    3.7&  275&    5.2,    5.7&  -& -& 1.22&  -& -\\
A1060             & $Z$       &103410.1-271402&   46&    3.7&   80&    5.5&  -& -& 1.22&  -& -\\
A1069             & $C$       &1037 6.8-081544&   42&    3.4,    1.9&  119&    7.6,    6.1&  218&   10.2&  .72&   72&    6.3,    4.6\\
A1146             & $C$       &105850.0-222806&   49&   20.6,   17.8&   85&   29.0,   25.9&  122&   35.9& 1.86&  148&   38.5,   43.2\\
A1314             & $M$(Red)  &113144.8 491918&   55&    7.2,    6.7&  -& -&  -& -&  .55&   64&    7.9,    7.4\\
A1631             & $C$       &125020.6-150453&  133&   12.9,    4.6&  317&   19.1,   10.0&  551&   23.3& 1.40&  511&   22.3,   14.2\\
A1644             & $C$       &125444.8-170834&  142&    7.7,    7.9&  313&   12.3,   11.8&  513&   17.6& 1.52&  524&   17.9,   18.0\\
A1656             & $M$(Blue) &125712.4+281323&  591&   10.1,   10.2& 1354&   16.2,   17.9& 2220&   22.6& 1.64&  -& -\\
A1656$^b$         & $M$(Blue) &125712.4+281323&   68&   16.9,   16.6&  114&   28.3,   28.1&  151&   37.3& 1.64&  156&   38.5,   38.5\\
A1656             & $Z$       &125712.4+281323&   68&   17.0&  113&   28.8&  150&   38.6& 1.64&  154&   39.7\\
A2040             & $M$(Red)  &151019.3+073633&  103&    7.9,    5.6&  -& -&  -& -&  .92&  -& -\\
A2151             & $M$(Red)  &160255.7+175332&   28&    7.0,    7.0&   60&   13.1,   13.1&   79&   16.1& 1.50&   79&   19.1,   19.7\\
A2151             & $Z$       &160255.7+175332&   27&    7.1&   59&   13.9&   72&   16.5& 1.50&   72&   16.5\\
A2197             & $M$(Blue) &162806.4+405707&    9&    4.5,    4.7&   22&   11.2,   11.9&  -& -& 1.22&  -& -\\
A2197             & $Z$       &162806.4+405707&    9&    4.7&   18&    9.8&  -& -& 1.22&  -& -\\
A2256             & $M$(Red)  &170627.0+784227&   62&   14.5,   14.1&  103&   20.8,   21.0&  138&   31.1& 2.70&  -& -\\
A2256             & $Z$       &170627.0+784227&   50&   12.8&   82&   19.0&   85&   19.9& 2.70&  -& -\\
A2353             & $C$       &213147.4-014827&   27&    6.6,    7.5&   44&    7.9,   11.2&   70&    9.2& 1.19&   47&    6.7,   11.4\\
A2362             & $C$       &213617.5-143242&   40&    1.4,    2.0&  119&    2.9,    4.6&  199&    2.4&  .66&   55&    1.6,    2.3\\
A2401             & $C$       &215534.6-202015&   73&    4.2,    4.2&  168&    6.0,    7.4&  285&    7.2&  .79&  132&    5.7,    6.3\\
A2426             & $C$       &221132.6-102555&   37&    5.3,    2.3&  117&    9.7,    3.8&  209&   14.6&  .66&   62&    6.8,    2.4\\
A2500             & $C$       &225108.6-254700&   31&    2.8,    1.9&   69&    4.2,    3.0&  125&    6.1&  .95&   63&    3.9,    2.6\\
A2554             & $C$       &230933.2-214708&   36&    7.5,    7.0&   93&   15.3,   13.9&  159&   23.2& 1.68&  181&   24.5,   26.7\\
A2554             & $M$(Blue) &230933.2-214708&   74&    8.6,    8.2&  198&   18.3,   16.9&  -& -& 1.68&  -& -\\
A2569             & $C$       &231521.1-130258&   38&    3.5,    4.0&   79&    6.2,    8.7&  118&    5.1&  .98&   79&    6.3,    8.7\\
A2593             & $M$(Red)  &232201.5+142305&   62&    4.7,    5.1&  133&    8.8,   10.4&  -& -& 1.40&  -& -\\
A2644             & $C$       &233802.0-001248&   29&    1.5,    1.2&   67&    2.1,    2.1&  136&    3.0&  .36&   15&     .7,     .5\\
A2670             & $C$       &235140.5-104146&   76&    6.9,    6.4&  159&   12.6,   13.1&  214&   12.3& 1.70&  245&   12.3,   17.2\\
A2670             & $M$(Blue) &235140.5-104146&   68&    8.5,    7.8&  150&   16.1,   16.6&  -& -& 1.70&  -& -\\
A2670             & $M$(Red)  &235140.5-104146&   68&   10.1,    8.9&  150&   19.9,   18.9&  -& -& 1.70&  -& -\\
\hline
\end{tabular}

%% file: tab2b.tex
%
%

\begin{tabular}{lcrrcrcrrrrc}
\hline \hline
\multicolumn{1}{c}{Name}
&\multicolumn{1}{c}{Sample}
&\multicolumn{1}{c}{Center}
&\multicolumn{1}{c}{$N(0.5)$ }   
&\multicolumn{1}{c}{$L_{B_j}(0.5)^a$}  
&\multicolumn{1}{c}{$N(1.0)$ }   
&\multicolumn{1}{c}{$L_{B_j}(1.0)^a$}  
&\multicolumn{1}{c}{$N(1.5)$ }   
&\multicolumn{1}{c}{$L_{B_j}(1.5)^a$}  
&\multicolumn{1}{c}{$R_{vir}$ }   
&\multicolumn{1}{c}{$N(R_{vir})$ }   
&\multicolumn{1}{c}{$L_{B_j}(R_{vir})^a$}
\\
\multicolumn{1}{c}{}
&\multicolumn{1}{c}{}
&\multicolumn{1}{c}{$\alpha(1950)-\delta(1950)$}
&\multicolumn{1}{c}{}   
&\multicolumn{1}{c}{$10^{11} h^{-2} L_{\odot}$}  
&\multicolumn{1}{c}{}   
&\multicolumn{1}{c}{$10^{11} h^{-2} L_{\odot}$}  
&\multicolumn{1}{c}{}   
&\multicolumn{1}{c}{$10^{11} h^{-2} L_{\odot}$}  
&\multicolumn{1}{c}{$h^{-1}\ Mpc$}  
&\multicolumn{1}{c}{}   
&\multicolumn{1}{c}{$10^{11} h^{-2} L_{\odot}$}  
\\
\hline
A2670             & $Z$       &235140.5-104146&   49&    6.3&  106&   13.0&  -& -& 1.70&  -& -\\
A2715             & $C$       &000011.6-345614&   20&    3.1,    2.9&   50&    6.4,    4.3&   91&   10.2&  .93&   46&    6.2,    4.2\\
A2717             & $C$       &240037.3-361239&   66&    2.2,    2.5&  156&    3.1,    4.7&  319&    6.3& 1.08&  172&    3.2,    4.8\\
A2717             & $M$(Blue) &000037.3-361239&   70&    4.5,    4.3&  154&    8.0,    8.3&  -& -& 1.08&  168&    9.6,    9.4\\
A2721             & $C$       &000335.0-345944&   38&    6.1,    6.3&   90&   13.6,   16.4&  138&   17.5& 1.61&  157&   19.7,   27.7\\
A2721             & $M$(Blue) &000335.0-345944&   69&   10.2,   10.6&  161&   22.3,   25.0&  -& -& 1.61&  -& -\\
A2734             & $C$       &000854.4-290659&   51&    3.2,    3.8&  131&    4.2,    6.8&  225&    5.3& 1.26&  184&    4.6,    8.1\\
A2755             & $C$       &001506.4-352739&   49&    6.3,    5.5&  105&   10.1,    9.0&  186&   14.9& 1.54&  197&   15.9,   15.4\\
A2798             & $C$       &003505.0-284835&   38&    7.1,    6.1&   71&   10.4,   11.5&  117&   21.3& 1.42&  108&   20.8,   10.5\\
A2799             & $C$       &003457.8-392420&   54&    2.5,    3.0&  126&    4.8,    7.0&  225&    7.2&  .84&   95&    3.8,    5.4\\
A2800             & $C$       &003535.4-252240&   39&    2.6,    3.0&   95&    4.4,    5.6&  175&    5.0&  .81&   70&    3.5,    4.3\\
A2877             & $C$       &010734.9-4613 1&  234&    5.8,    6.2&  530&    7.0,    8.7& 1015&    9.9& 1.77& 1379&   10.9,   16.5\\
A2911             & $C$       &012349.1-381145&   40&    3.9,    3.8&   95&   11.4,    5.6&  162&   19.7& 1.09&  106&   11.5,    6.0\\
A3093             & $C$       &030917.6-473545&   39&    3.8,    3.8&   80&    4.6,    5.6&  189&   11.4&  .88&   71&    4.6,    5.3\\
A3094             & $C$       &030932.3-271027&   66&    4.2,    4.7&  164&    9.2,    9.9&  257&   10.5& 1.31&  223&   10.5,   12.3\\
A3111             & $C$       &031557.1-455754&   59&    4.9,    1.4&  130&   10.8,    5.2&  228&   14.7&  .32&   30&    3.1,    1.0\\
A3122             & $C$       &032022.4-412948&   66&    2.8,    3.5&  136&    4.9,    6.5&  216&    4.8& 1.55&  223&    4.6,    9.6\\
A3126             & $C$       &032716.9-555300&   61&    5.1,    5.0&  123&    8.2,    9.9&  195&   11.7& 2.11&  281&   11.0,   24.0\\
A3126             & $M$(Blue) &032716.9-555300&   35&    6.0,    5.7&   74&   10.7,   11.8&  -& -& 2.11&  -& -\\
A3128             & $C$       &032928.3-524035&  113&    8.9,    8.6&  285&   16.6,   16.0&  409&   19.2& 1.58&  434&   19.3,   21.9\\
A3128             & $M$(Blue) &032928.3-524035&   68&   10.3,    9.8&  158&   20.2,   18.5&  -& -& 1.58&  -& -\\
A3142             & $C$       &033455.4-395609&   43&    7.2,    4.5&   87&   10.5,    7.0&  133&   12.3& 1.47&  132&   12.6,   10.0\\
A3151             & $C$       &033830.7-285023&   61&    4.2,    1.6&  132&    5.0,    2.6&  250&   11.1&  .47&   57&    4.1,    1.6\\
A3158             & $C$       &034141.2-534728&  137&    6.3,    6.7&  285&   11.0,   12.4&  472&   17.2& 1.95&  669&   19.4,   26.0\\
A3194             & $C$       &035708.7-301916&   44&    5.2,    6.2&  109&   11.3,   14.9&  187&   16.4& 1.61&  203&   17.0,   26.2\\
A3223             & $C$       &040611.6-311045&   86&    4.3,    4.7&  192&    7.6,    8.2&  300&    9.1& 1.29&  263&    9.4,   10.8\\
A3266             & $C$       &043007.7-614031&  124&    6.4,    4.9&  297&   12.1,   10.4&  500&   19.0& 2.21&  780&   24.8,   28.9\\
A3334             & $C$       &051703.0-583632&   35&    4.2,    4.2&   68&    6.1,    8.3&  100&    6.1& 1.39&   96&    6.7,   11.9\\
A3334             & $M$(Blue) &051703.0-583632&   69&    5.5,    5.5&  148&    8.6,   11.1&  -& -& 1.39&  -& -\\
A3354             & $C$       &053245.2-284245&   52&    3.0,    2.9&  184&   10.2,    6.8&  331&   13.3&  .72&  102&    8.2,    5.4\\
A3360             & $C$       &053836.6-432531&   31&    4.5,    4.6&   77&    6.2,    8.6&  115&    8.4& 1.67&  140&    9.1,   17.1\\
A3360             & $M$(Blue) &053836.6-432531&   67&    5.0,    5.0&  157&    9.8,   11.9&  -& -& 1.67&  -& -\\
A3376             & $C$       &060037.8-395622&   73&    2.7,    3.0&  203&    5.3,    6.8&  309&    5.1& 1.38&  288&    5.4,    8.9\\
A3381             & $C$       &060807.6-333244&  110&    4.4,    1.7&  273&    6.6,    2.7&  537&   15.4&  .59&  126&    4.6,    2.0\\
A3391             & $C$       &062514.2-533953&   72&    4.9,    4.9&  189&   10.7,   10.2&  339&   18.1& 1.33&  278&   16.3,   16.4\\
A3395             & $C$       &062634.7-542433&  113&    5.7,    4.8&  261&   10.6,    8.9&  425&   15.1& 1.70&  496&   16.0,   15.9\\
A3532             & $C$       &125432.5-300451&   91&    6.4,    4.8&  252&   12.7,   10.6&  467&   18.8& 1.48&  463&   18.9,   17.3\\
A3556             & $C$       &132130.3-312641&   93&    4.7,    4.9&  251&   10.5,    9.6&  547&   24.4& 1.28&  409&   16.2,   15.0\\
A3558             & $C$       &132507.1-311347&  171&    8.7,    8.8&  467&   23.5,   24.0&  805&   35.6& 1.95& 1151&   50.3,   55.5\\
A3558             & $M$(Blue) &132507.1-311347&  369&    8.3,    8.8&  -& -&  -& -& 1.95&  -& -\\
A3558             & $Z$       &132507.1-311347&   53&    9.7&  -& -&  -& -& 1.95&  -& -\\
A3559             & $C$       &132723.9-291830&   91&    6.0,    6.3&  248&   10.7,   11.7&  489&   20.4&  .91&  215&    9.5,   10.3\\
A3571             & $C$       &134428.3-323710&  255&   10.4,   10.9&  644&   19.3,   21.1& 1087&   28.5& 2.09& 1746&   36.0,   45.7\\
A3574             & $C$       &134558.3-301238&  433&    6.1,    5.8& 1337&    8.5,    9.3& 2595&   11.3&  .98& 1303&    8.6,    9.2\\
A3574             & $M$(Blue) &134558.3-301238&   38&    4.2,    3.9&  -& -&  -& -&  .98&  -& -\\
A3574             & $Z$       &134558.3-301238&   21&    3.7&  -& -&  -& -&  .98&  -& -\\
A3651             & $C$       &194826.9-551552&  111&    9.1,    9.4&  264&   16.6,   18.1&  428&   24.4& 1.25&  348&   21.3,   19.7\\
A3667             & $C$       &200825.8-565733&  131&    9.8,    9.5&  371&   21.4,   22.2&  646&   29.6& 1.94&  882&   40.8,   40.1\\
A3667             & $M$(Blue) &200825.8-565733&   43&    6.3,    6.0&  110&   15.1,   15.4&  158&   19.2& 1.94&  -& -\\
A3667             & $Z$       &200825.8-565733&   19&    5.6&   53&   14.7&   73&   19.4& 1.94&  -& -\\
A3693             & $C$       &203110.2-344325&   42&    4.3,    1.9&   90&    8.1,    4.8&  166&   13.4&  .96&   89&    8.3,    4.6\\
A3695             & $C$       &203133.1-355824&   60&    9.5,   10.3&  136&   17.2,   18.7&  210&   20.8& 1.56&  214&   20.6,   26.6\\
A3705             & $C$       &203900.1-352256&   73&    9.1,    8.7&  174&   19.7,   18.7&  288&   28.7& 1.75&  327&   29.6,   32.4\\
A3705             & $M$(Blue) &203900.1-352256&   70&    7.8,    7.5&  169&   15.0,   15.1&  -& -& 1.75&  -& -\\
A3733             & $C$       &205836.0-281416&   83&    2.9,    3.3&  240&    5.5,    7.4&  414&    5.2& 1.22&  301&    5.3,    8.3\\
A3744             & $C$       &210427.9-253800&  107&    3.7,    4.3&  330&    6.5,    7.7&  597&    8.0& 1.02&  340&    6.6,    7.8\\
\hline
\end{tabular}

%% file: tab2c.tex
%
%

\begin{tabular}{lcrrcrcrrrrc}
\hline \hline
\multicolumn{1}{c}{Name}
&\multicolumn{1}{c}{Sample}
&\multicolumn{1}{c}{Center}
&\multicolumn{1}{c}{$N(0.5)$ }   
&\multicolumn{1}{c}{$L_{B_j}(0.5)^a$}  
&\multicolumn{1}{c}{$N(1.0)$ }   
&\multicolumn{1}{c}{$L_{B_j}(1.0)^a$}  
&\multicolumn{1}{c}{$N(1.5)$ }   
&\multicolumn{1}{c}{$L_{B_j}(1.5)^a$}  
&\multicolumn{1}{c}{$R_{vir}$ }   
&\multicolumn{1}{c}{$N(R_{vir})$ }   
&\multicolumn{1}{c}{$L_{B_j}(R_{vir})^a$}
\\
\multicolumn{1}{c}{}
&\multicolumn{1}{c}{}
&\multicolumn{1}{c}{$\alpha(1950)-\delta(1950)$}
&\multicolumn{1}{c}{}   
&\multicolumn{1}{c}{$10^{11} h^{-2} L_{\odot}$}  
&\multicolumn{1}{c}{}   
&\multicolumn{1}{c}{$10^{11} h^{-2} L_{\odot}$}  
&\multicolumn{1}{c}{}   
&\multicolumn{1}{c}{$10^{11} h^{-2} L_{\odot}$}  
&\multicolumn{1}{c}{$h^{-1}\ Mpc$}  
&\multicolumn{1}{c}{}   
&\multicolumn{1}{c}{$10^{11} h^{-2} L_{\odot}$}  
\\
\hline
A3809             & $C$       &214405.9-440914&   63&    3.3,    3.6&  172&    6.1,    7.7&  293&    7.1&  .96&  162&    5.8,    7.1\\
A3822             & $C$       &215043.8-580510&  120&    9.3,    8.0&  329&   24.8,   22.2&  506&   34.2& 1.62&  545&   36.3,   33.8\\
A3825             & $C$       &215444.4-603244&   80&    5.5,    5.0&  186&   11.9,   11.3&  328&   18.6& 1.40&  298&   17.0,   16.6\\
A3879             & $C$       &222414.3-6911 3&   49&    3.8,    4.1&   97&    4.7,    5.8&  170&    5.2&  .80&   81&    4.9,    5.2\\
A3880             & $C$       &222502.0-304936&   64&    3.0,    3.2&  151&    6.2,    7.9&  299&   13.7& 1.65&  365&   14.7,   19.7\\
A3921             & $C$       &224645.3-643942&   47&    6.9,    7.2&  118&   12.0,   13.5&  186&   14.8&  .98&  117&   12.1,   13.4\\
A4008             & $C$       &232739.1-393210&   59&    3.4,    3.4&  156&    5.1,    6.2&  275&    6.8&  .85&  126&    4.7,    5.3\\
A4010             & $C$       &232851.7-364643&   37&    4.7,    5.6&   82&    7.9,    9.0&  139&   13.7& 1.25&  102&    8.7,   10.6\\
A4053             & $C$       &235207.7-275757&   43&    3.5,    2.2&  109&    6.0,    3.9&  178&    7.7& 1.23&  136&    6.2,    4.6\\
A4067             & $C$       &235621.3-605412&   33&    3.6,    3.7&   84&    8.1,    8.7&  134&    9.0& 1.00&   84&    8.1,    8.7\\
S84               & $C$       &004659.4-294712&   35&    5.4,    3.5&   71&    8.1,    6.9&  105&    8.9&  .66&   50&    6.8,    5.1\\
S373              & $C$       &033408.4-352339& 2631&    1.7,    3.2&10793& *,    7.8&22620& *&  .62& 4150&    1.3,    3.7\\
S463              & $C$       &042802.5-535644&  151&    9.4,    9.8&  327&   14.6,   14.2&  566&   18.8& 1.22&  416&   15.8,   16.2\\
S721              & $C$       &130316.8-372100&  123&    4.0,    3.4&  298&    7.3,    7.5&  524&   12.7& 1.38&  463&   11.4,   12.2\\
S753              & $C$       &140019.9-334851&  423&    2.9,    2.8& 1451&    5.0,    5.3& 3299&    6.1& 1.07& 1653&    5.1,    5.7\\
S753              & $M$(Blue) &140019.9-334851&   33&    2.6,    2.3&  -& -&  -& -& 1.07&  -& -\\
S753              & $Z$       &140019.9-334851&   17&    2.6&  -& -&  -& -& 1.07&  -& -\\
S805              & $C$       &184802.7-631810&  402&    1.6,    2.0& 1843&    5.0,    7.5& 4149&   10.0& 1.08& 2133&    6.3,    7.6\\
S987              & $C$       &215907.1-223859&   43&    5.5,    6.1&  121&   13.1,   15.9&  232&   21.8& 1.35&  206&   21.4,   25.4\\
S1157             & $C$       &234903.9-344355&   43&    4.7,    2.6&  152&    8.3,    4.7&  274&   12.0& 1.16&  191&    8.9,    5.4\\
S1157             & $M$(Blue) &234903.9-344355&   46&    5.1,    2.8&  151&   11.1,    5.8&  -& -& 1.16&  -& -\\
AWM1              & $M$(Blue) &091411.5 201434&   11&    5.1,    5.0&   19&    8.3,    8.7&  -& -&  .88&   19&    8.4,    8.7\\
AWM1              & $Z$       &091411.5 201434&   11&    5.1&   19&    8.8&  -& -&  .88&   19&    8.8\\
AWM4              & $M$(Red)  &160247.7+240437&   38&    1.5,     .9&  -& -&  -& -&  .24&   20&    1.3,     .6\\
AWM4              & $Z$       &160247.7+240437&    7&    1.4&  -& -&  -& -&  .24&  -& -\\
AWM7              & $M$(Blue) &025122.7+412330&   14&    3.0,    2.6&  -& -&  -& -& 1.73&  -& -\\
AWM7              & $Z$       &025122.7+412330&   14&    3.2&  -& -&  -& -& 1.73&  -& -\\
DC0003-50         & $C$       &000332.2-505524&   85&    2.3,    2.7&  254&    3.4,    4.9&  477&    4.3&  .70&  146&    3.2,    4.0\\
Eridanus          & $C$       &033759.4-184707& 1386&     .5,     .9& 5270& *,   15.0&11730& *&  .53& 1536&     .5,    1.0\\
MKW1             & $C$        &095812.1-024317&  208&    1.5,    1.9&  676&    2.2,    3.8& 1495&    3.4&  .45&  185&    1.5,    1.8\\
MKW4              & $M$(Red)  &120146.5+020748&   47&    2.4,    2.4&  -& -&  -& -& 1.05&  -& -\\
MKW4              & $Z$       &120146.5+020748&   24&    2.8&  -& -&  -& -& 1.05&  -& -\\
Ursa Major  & $Z$       &115454.1+511207&   23&     .5&  -& -&  -& -&  .26&    8&     .2\\
Virgo             & $M$(Blue) &122332.4+130259&  145&    1.8,    2.8&  -& -&  -& -& 1.26&  -& -\\
Virgo             & $Z$       &122332.4+130259&  101&    3.5&  -& -&  -& -& 1.26&  -& -\\
\hline
\end{tabular}

%% file: comm_tab2.tex
{\footnotesize\parindent=3mm
$^a$ For $C$-- and $M$--clusters
we report both $L_{B_j,c}$ and $L_{B_j,f}$ 
within 0.5 and 1.0 \h, and $R_{vir}$; we report 
only $L_{B_j,c}$ within 1.5. 
The asterisk,
$*$, denotes luminosity estimates which result
negative values; this happens for two poor clusters (S753 and Eridanus)
whose local galaxy field is much poorer than the mean field.\\
$^b$ The sample  of A1656 in $B_{Zwicky}$--magnitude band (cf. Table~1).}

%% file: tab3a.tex
%
%
  
\begin{tabular}{lccc}
\hline \hline
\multicolumn{1}{c}{Name}
&\multicolumn{1}{c}{$M/L_{B_j}$ (range)}
&\multicolumn{1}{c}{$M/L_{B_j,c}$ ($C$--clusters)}   
&\multicolumn{1}{c}{$M/L_{B_j,f}$ ($C$--clusters)}   
\\
\multicolumn{1}{c}{}
&\multicolumn{1}{c}{$h\ (M_{\odot}/L_{\odot})$}  
&\multicolumn{1}{c}{$h\ (M_{\odot}/L_{\odot})$}  
&\multicolumn{1}{c}{$h\ (M_{\odot}/L_{\odot})$}  
\\
\hline
A85               & 246-917   &   481&   422\\
A119              & 143-313   &   181&   143\\
A194              & 89-377    &   239&   198\\
A229              & 173-   412&   295&   246\\
A256              &   147-   428&   156&   205\\
A295              &    71-   207&   130&   133\\
A420              &   249-   365&   279&   317\\
A426              & 280-458   &   -&-\\
A458              & 296-480   &   443&   309\\
A496              &   288-   808&   416&   288\\
A514              &   500-  1131&   687&   500\\
A524              &    91-   230&    91&   170\\
A539              & 110-482   &   -&-\\
A576              & 588-736   &   -&-\\
A978              &   326-   619&   424&   326\\
A1060             & 162-327   &   216&   162\\
A1069             &   121-   298&   121&   164\\
A1146             &   171-   247&   191&   171\\
A1314             & 40-43     &    -&-\\
A1631             &   149-   415&   227&   356\\
A1644             &   219-   285&   251&   251\\
A1656             & 121-225   &   -&-\\
A2040             & 128-181   &   -&-\\
A2151             & 287-351   &   -&-\\
A2197             & 257-385   &   -&-\\
A2256             & 560-876   &   -&-\\
A2353             &   170-   309&   309&   180\\
A2362             &   158-   398&   398&   269\\
A2401             &   145-   255&   171&   154\\
A2426             &    58-   202&    70&   202\\
A2500             &   204-   396&   260&   383\\
A2554             & 228-529   &   247&   228\\
A2569             &   230-   510&   315&   230\\
A2593             & 173-283   &   -&-\\
A2644             &    62-   156&   110&   156\\
A2670             & 215-451   &   451&   322\\
A2715             &   241-   421&   274&   408\\
A2717             & 154-446   &   466&   308\\
A2721             & 184-454   &   280&   200\\
A2734             &   413-   723&   723&   413\\
A2755             &   387-   500&   412&   426\\
A2798             &    90-   229&    91&   180\\
A2799             &   200-   312&   312&   221\\
A2800             &   184-   382&   269&   222\\
A2877             &   297-   639&   449&   297\\
A2911             &    95-   272&   139&   265\\
A3093             &   116-   197&   175&   153\\
A3094             &   316-   514&   443&   380\\
A3111             &    11-    53&    17&    53\\
A3122             &   647-  1340&  1340&   647\\
A3126             & 325-1080  &   711&   325\\
A3128             & 261-369   &   369&   325\\
A3142             &   345-   594&   423&   531\\
A3151             &    40-   131&    48&   125\\
A3158             &   360-   660&   483&   360\\
A3194             &   280-   552&   432&   280\\
\hline
\end{tabular}

%% file: tab3b.tex
%
%
  
\begin{tabular}{lccc}
\hline \hline
\multicolumn{1}{c}{Name}
&\multicolumn{1}{c}{$M/L_{B_j}$ (range)}
&\multicolumn{1}{c}{$M/L_{B_j,c}$ ($C$--clusters)}   
&\multicolumn{1}{c}{$M/L_{B_j,f}$ ($C$--clusters)}   
\\
\multicolumn{1}{c}{}
&\multicolumn{1}{c}{$h\ (M_{\odot}/L_{\odot})$}  
&\multicolumn{1}{c}{$h\ (M_{\odot}/L_{\odot})$}  
&\multicolumn{1}{c}{$h\ (M_{\odot}/L_{\odot})$}  
\\
\hline
A3223             &   301-   397&   348&   301\\
A3266             &   405-   547&   473&   405\\
A3334             & 282-650    &   568&   320\\
A3354             & 71-156    &   71 & 107\\
A3360             &   461-  1007&  1007&   534\\
A3376             &   407-   766&   669&   407\\
A3381             &    45-   200&    81&   186\\
A3391             &   217-   366&   222&   221\\
A3395             &   355-   762&   355&   359\\
A3532             &   172-   251&   172&   188\\
A3556             &   165-   322&   216&   233\\
A3558             & 208-507   &   229&   208\\
A3559             &    94-   134&   134&   123\\
A3571             &   179-   500&   227&   179\\
A3574             & 151-256   &   165&   153\\
A3651             &   171-   200&   184&   200\\
A3667             & 277-551   &   288&   293\\
A3693             &   100-   235&   103&   186\\
A3695             &   128-   187&   187&   145\\
A3705             & 232-439   &   254&   232\\
A3733             &   329-   666&   545&   350\\
A3744             &   161-   203&   191&   161\\
A3809             &   223-   362&   283&   229\\
A3822             &   109-   127&   116&   124\\
A3825             &   218-   354&   229&   234\\
A3879             &   113-   216&   140&   132\\
A3880             &   294-   938&   396&   294\\
A3921             &   151-   198&   168&   151\\
A4008             &   187-   233&   220&   194\\
A4010             &   212-   325&   308&   253\\
A4053             &   298-   495&   334&   448\\
A4067             &   122-   168&   131&   122\\
S84               &    55-    90&    55&    74\\
S373              &    58-   242&   242&    84\\
S463              &   103-   126&   126&   123\\
S721              &   251-   596&   274&   254\\
S753              & 232-317   &   257&   232\\
S805              &   143-   336&   185&   153\\
S987              &   126-   393&   150&   126\\
S1157             & 249-602   &   363&   602\\
AWM1              & 127-148   &   -&-\\
AWM4              & 17-41     &   -&-\\
AWM7              & 917-1104  &   -&-\\
DC0003-50         &   142-   286&   178&   142\\
Eridanus          &    50-   776&   776&   417\\
MKW1              &    74-   127&   109&    91\\
MKW4              & 319-367   &   -&-\\
Ursa Major              & 164-252   &   -&-\\
Virgo             & 242-459   &   -&-\\
\hline
\end{tabular}